\documentclass[12pt]{article}

\usepackage{scicite}
\usepackage{times}

\usepackage{amssymb,graphicx,xcolor,hyperref,pdfpages}

\newcommand{\ea}{{\it et al.}}

\newcommand{\kms}{km~$\rm{s}^{-1}$}

\newcommand{\C}{$^{12}$C}
\newcommand{\OX}{$^{16}$O}
\newcommand{\cms}{cm~s$^{-1}$}
\newcommand{\ms}{m~s$^{-1}$}

\newcommand{\beq}{\begin{equation}}
\newcommand{\eeq}{\end{equation}}
\newcommand{\bdm}{\begin{displaymath}}
\newcommand{\edm}{\end{displaymath}}

\newcommand{\athena}{{\sc Athena-RFX}}

\newcommand{\revi}[0]{\textcolor{black}}
\newcommand{\revii}[0]{\textcolor{black}}
\newcommand{\reviii}[0]{\textcolor{black}}

\newenvironment{sciabstract}{%
\begin{quote} \bf}
{\end{quote}}

\title{A Unified Mechanism for Unconfined Deflagration-to-Detonation
Transition in Terrestrial Chemical Systems and Type Ia Supernovae}

\author{
Alexei Y. Poludnenko,$^{1,2\ast}$ Jessica Chambers,$^{3}$ Kareem Ahmed,$^{3}$\\
Vadim N. Gamezo,$^{4}$ Brian D. Taylor$^{5}$\\
\\
\normalsize{$^{1}$Department of Aerospace Engineering}\\ \normalsize{Texas A\&M University, College Station, TX 77843, USA}\\
\normalsize{$^{2}$Department of Mechanical Engineering}\\ \normalsize{University
of Connecticut, Storrs, CT 06269, USA}\\
\normalsize{$^{3}$Department of Mechanical and Aerospace Engineering}\\ \normalsize{University of Central Florida, Orlando, FL 32816, USA}\\
\normalsize{$^{4}$Laboratories for Computational Physics and Fluid Dynamics}\\ \normalsize{Naval Research Laboratory, Washington, DC 20745, USA}\\
\normalsize{$^{5}$Munitions Directorate}\\ \normalsize{Air Force Research Laboratory, Eglin, FL 32542, USA}\\
\\
\normalsize{$^\ast$To whom correspondence should be addressed; E-mail:  alexei.poludnenko@uconn.edu}
}

\date{}

\begin{document}

\maketitle 

A general theory of the deflagration-to-detonation transition is presented
along with its experimental and numerical confirmation, as well as its
application to Type Ia supernovae.

\begin{sciabstract}

The nature of type Ia supernovae (SNIa) - thermonuclear explosions of white
dwarf stars - is an open question in astrophysics. Virtually all existing
theoretical models of normal, bright SNIa require the explosion to produce a
detonation in order to consume all of stellar material, but the mechanism for
the deflagration-to-detonation transition (DDT) remains unclear. We present a
unified theory of turbulence-induced DDT that describes the mechanism and
conditions for initiating detonation both in unconfined chemical and
thermonuclear explosions. The model is validated using experiments with chemical
flames and numerical simulations of thermonuclear flames. We use the developed
theory to determine criteria for detonation initiation in the single-degenerate
Chandrasekhar-mass SNIa model, and show that DDT is almost inevitable at
densities $10^7 - 10^8$ g~cm$^{-3}$.

\end{sciabstract}

Type Ia supernovae (SNIa) are an important tool for measuring cosmological
distances. Techniques used to calibrate SNIa for this purpose rely on empirical
correlations \cite{Phillips_1993, Riess_1995, Jha_2007} due to incomplete
theoretical understanding of the mechanisms responsible for SNIa. Theoretical
models of SNIa have remained limited due to uncertainties in the explosion
mechanisms.

Explosions can involve two distinct types of combustion waves that differ by the
propagation mechanism: deflagrations and detonations. Deflagrations, or flames,
are relatively slow and propagate subsonically through heat conduction,
diffusion, and advection. Detonations move supersonically due to ignition by
shock compression, and therefore always involve strong shocks and propagate with
the shock speed.

SNIa explosions are driven by fast thermonuclear burning in \C/\OX~ white dwarf
(WD) stars with a mass close to, or below, the Chandrasekhar-mass limit of
$\approx\!1.4$ solar masses \cite{Hoyle_1960} - the maximum mass of a WD
supported against the gravitational collapse by the electron degeneracy
pressure. Beyond this general statement, however, the exact mechanisms of SNIa
remain unclear \cite{Wang_2012, Hillebrandt_2013, Maoz_2014, Ropke_2018}, with a
number of possible scenarios. These include the classical Chandrasekhar-mass
(M$_{ch}$) model \cite{Khokhlov_1991, Hoflich_1995, Gamezo_2003,
Seitenzahl_2013}, sub-Chandrasekhar-mass (sub-M$_{ch}$) models
\cite{Woosley_1994, Hoflich_1996, Ruiter_2011, Goldstein_2018},
double-degenerate (DD) models that include mergers or collisions of two WDs
\cite{Mochkovitch_1989, Guillochon_2010, Pakmor_2010, Papish_2016}, and
gravitationally confined detonations \cite{Plewa_2004, Seitenzahl_2016}.

All SNIa explosion scenarios share a common characteristic: to produce a normal,
bright SNIa, detonation must be triggered in the stellar interior at some stage
of the explosion. Such supersonic combustion is required to consume the entire
WD, including its outer layers, which expand supersonically as the material
becomes gravitationally unbound. Observations place tight upper limits on the
amount of unburned $^{12}$C in the resulting SNIa ejecta indicating nearly
complete combustion of the stellar material in the explosion \cite{Marion_2009,
Parrent_2014}. Pure deflagration M$_{ch}$ explosion scenarios have been
previously suggested as the mechanism forming subluminous classes of supernovae,
e.g., Type Iax \cite{Fink_2013}. 

Generally, it is more difficult to ignite a detonation than a flame in a
chemically or thermonuclearly explosive mixture. Detonation can typically form
in one of three ways: i) strong ignition, in which a detonation is directly
initiated by a pre-existing, or externally introduced, strong shock
\cite{Meyer_Oppenheim_1971}; ii) weak ignition, in which a detonation develops
via a spontaneous reaction wave propagating through a gradient of temperature or
composition \cite{Zeldovich_1970, Lee_1978} created by weaker shocks or other
processes (gradient mechanism); and 3) deflagration-to-detonation transition
(DDT), during which shocks are produced by fast flames and the detonation may
sometimes form at the final stages via the gradient mechanism \cite{Oran_2007}.

While strong ignition may be realized in the gravitationally confined model
\cite{Plewa_2004, Seitenzahl_2016} or in the case of WD collisions
\cite{Raskin_2010, Papish_2016}, more widely accepted models, such as the
M$_{ch}$ and sub-M$_{ch}$ scenarios, as well as the DD mergers, generally do not
have any natural mechanism to form pre-existing shocks of sufficient strength to
directly ignite a detonation. For example, in the classical M$_{ch}$ scenario,
detonation formation requires a DDT because burning must initially propagate via
subsonic flames to pre-expand a star and produce intermediate-mass elements
\cite{Khokhlov_1991, Gamezo_2005, Ropke_Niemeyer_2007}. In the sub-M$_{ch}$,
both the gradient and the DDT mechanisms of detonation formation cannot be ruled
out. Material compression produced by accretion in the surface layers leads to
ignition, which could produce either: localized hotspots directly triggering a
detonation through the gradient mechanism; or a flame, which could subsequently
trigger a DDT. We focus on the DDT in degenerate stellar matter as the potential
mechanism of detonation ignition in the two leading SNIa scenarios, namely
M$_{ch}$ and sub-M$_{ch}$. Although, DDT is generally not considered in the
context of the DD models, the detailed ignition process in that scenario remains
unclear.

Thermonuclear combustion waves are qualitatively similar to chemical combustion
waves on Earth, as they  are controlled by the same physical mechanisms. This
similarity allows us to seek insights into the fundamental aspects of the
physical processes controlling SNIa explosions using theoretical, numerical, and
experimental results obtained for terrestrial chemical systems. DDT is also
relevant to terrestrial applications ranging from detonation-based propulsion
and power-generation systems, e.g., detonation engines \cite{Kailas_2000,
Kailas_2003, Wolanski_2013, Lu_2014, Rankin_2017}, to the industrial safety of
mining operations \cite{Zipf_2014}, fuel-storage, chemical processing
\cite{Tam_2016, Johnson_2017}, and  nuclear power-generation facilities
\cite{Luangdilok_2017}. Large-scale industrial accidents in Buncefield, UK
\cite{Tam_2011, Taveau_2012} (fuel-storage facility, 2005), Sago Mine, US
\cite{McMahon_2007} (coal mine, 2006), Jaipur, India \cite{Johnson_2012}
(chemical processing plant, 2009), and Fukushima, Japan \cite{Fukushima_2011}
(nuclear power plant, 2011) may all have involved a DDT.

The DDT process in SNIa occurs in essentially unconfined conditions, in the
sense that there are no walls or obstructions that are usually present in
terrestrial settings. The mechanism of such unconfined DDT is not understood in
chemical systems on Earth. Direct numerical simulations (DNS) have shown that
chemical flames interacting with high-intensity turbulence can spontaneously
accelerate and produce strong shocks or detonations in a completely unconfined
setting \cite{Poludnenko_2011}. We present an experimental confirmation of this
shock-generation and DDT mechanism in terrestrial systems, and apply it to SNIa.

\section*{Theory of turbulence-induced DDT}
\label{Theory}

DNS of chemical flames in high-intensity turbulent flows have shown that
turbulent flames with burning speeds $S_T$ above the speed of a Chapman-Jouguet
(CJ) deflagration, $S_{CJ}$, are intrinsically unstable and can transition to
detonations \cite{Poludnenko_2011}. The $S_{CJ}$ is defined as the flame speed,
at which the flow on the product side of the flame becomes sonic in the
reference frame co-moving with the flame. Consequently, the condition for the
turbulence-driven spontaneous DDT (tDDT) can be written as \cite{Poludnenko_2011}
\beq
S_T > S_{CJ} = c_s / \alpha,
\label{e:SCJ}
\eeq
where $c_s$ is the sound speed in hot products, and $\alpha = \rho_f / \rho_p$
is the ratio of the densities of fuel $\rho_f$ and combustion
products $\rho_p$.

\begin{figure}[t]
\centering
\includegraphics[clip, width=0.9\textwidth]{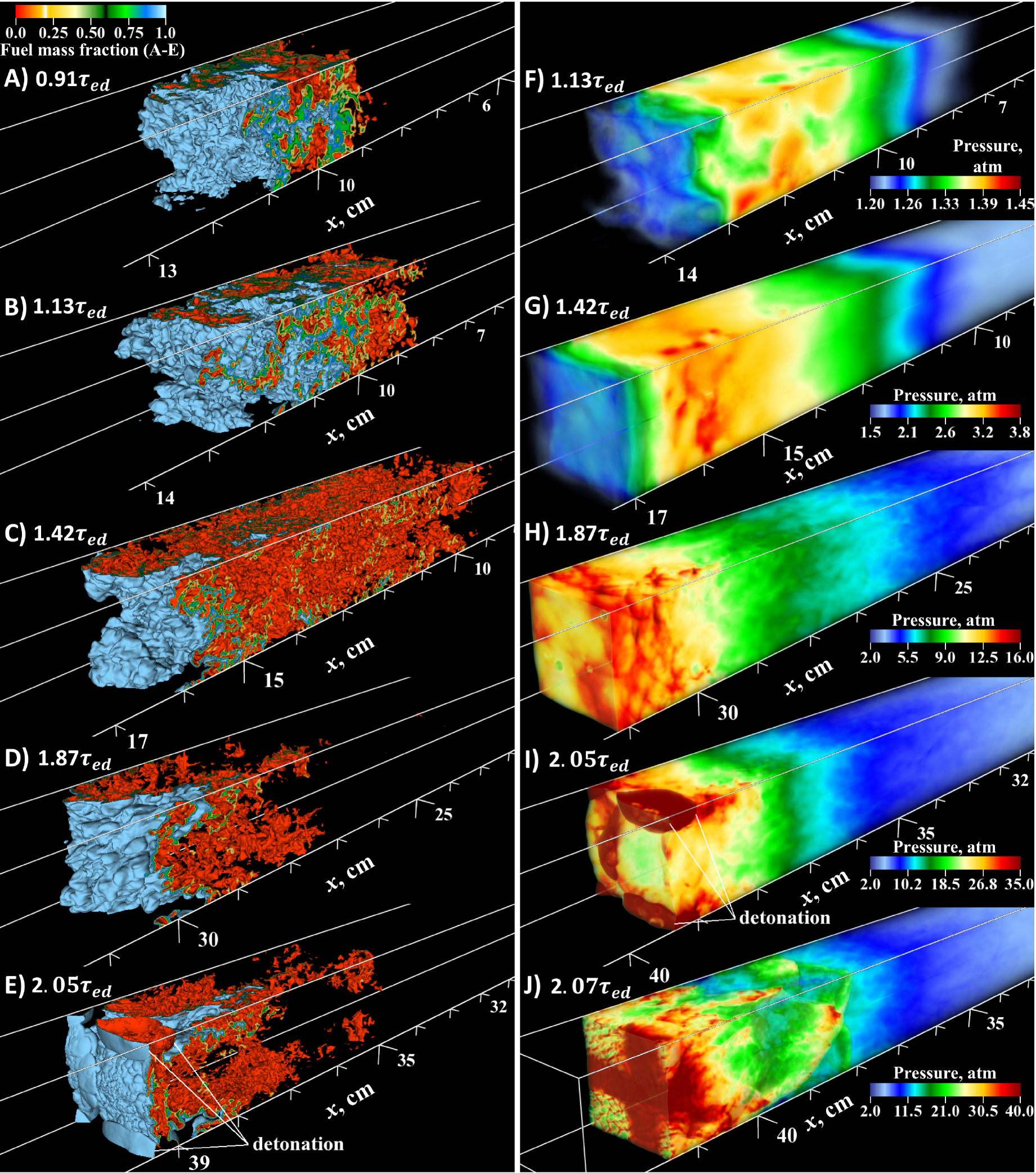}
\caption{
{\bf Evolution of a turbulent flame and transition to a detonation in a
stoichiometric methane-air mixture} (simulation reported as case 11 in
Ref.~\cite{Poludnenko_2011}). \textit{Panels A - E:} structure of a turbulent
flame propagating to the left. Shown is the isovolume bounded by the two
isosurfaces of the fuel mass fraction $Y = 0.05$ and $Y = 0.95$. Colors indicate
the value of $Y$. \textit{Panels F - J:} corresponding pressure fields. Color
scales in panels F - J show pressure normalized by the upstream pressure in the
domain, $P_0 =1$ atm, and the color scale changes between panels as the maximum
pressure increases. Time for each frame is given in units of the eddy turnover
time $\tau_{ed} = 0.367$ ms. Corresponding turbulent flame speed and maximum
pressures in the domain are shown in Fig.\ref{f:graphs}c. This simulation is
shown in Movie S1.}
\label{f:CH4}
\end{figure}

\begin{figure}[t]
\centering
\includegraphics[clip, width=1.0\textwidth]{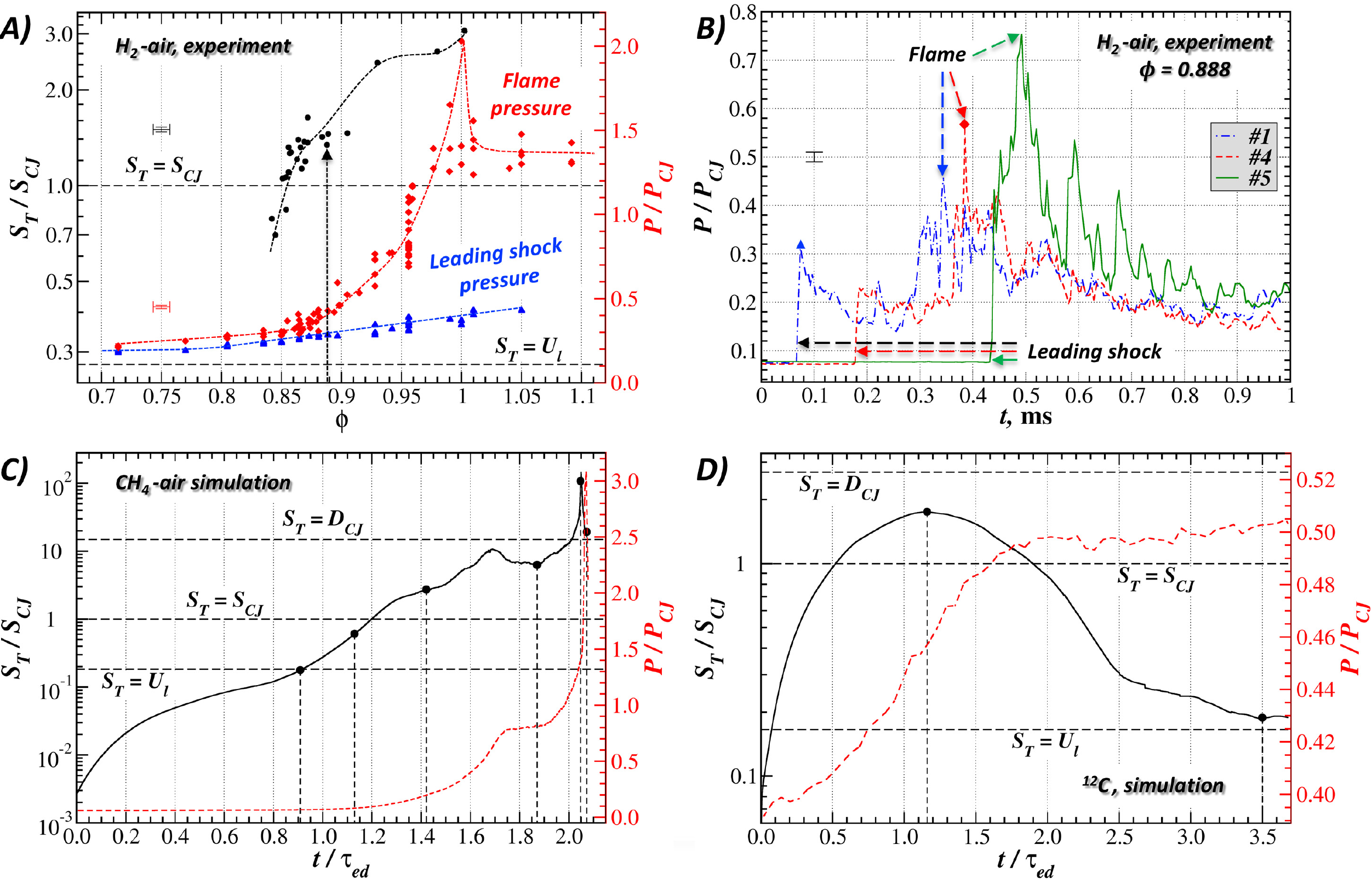}
\caption{
\textbf{Pressures, $P$, and turbulent flame speeds, $S_T$, observed in
experiments (A,B) and simulations (C,D) for chemical (A-C) and thermonuclear (D)
flames.} \textit{Panel A:} summary of all experiments in hydrogen-air mixtures
as a function of the equivalence ratio. Black circles: $S_T$ measured at the
beginning of the first test section (Fig.~\ref{f:facility}). Values of $S_T$ are
scaled by the CJ deflagration velocity $S_{CJ}$, which depends on $\phi$ and
varies between $192$ and $292$ m~s$^{-1}$. Blue triangles: peak pressures of the
leading shock. Red diamonds: maximum pressures generated by the turbulent flame
in the first test section. All pressures are scaled by the CJ detonation
pressure $P_{CJ}$, which depends on $\phi$ and varies between $14.0$ and $15.8$
atm. Blue triangles and red diamonds in panel A correspond to pressure peaks indicated by a
blue triangle and a red diamond in panel B, respectively. Black arrow marks the
experiment shown in panels F - H in Fig.~\ref{f:schlierens}, while the black
circle immediately to the right marks the experiment shown in panels I and J.
\textit{Panel B:} pressure histories recorded in the experiment with $\phi=
0.888$ by three pressure transducers indicated in Fig.~\ref{f:facility}.
Pressure is scaled by $P_{CJ} = 15.06$ atm. The first peak in each curve
corresponds to the leading shock. The second pressure maximum is associated with
the compressed region ahead of the flame (Fig.~\ref{f:schlierens}).
\textit{Panel C:} $S_T$ and maximum instantaneous $P$ in the computational
domain as functions of time in the same simulation as Fig.~\ref{f:CH4}. Initial
pressure $P_0 =1$ atm. $S_T$ is scaled by $S_{CJ}= 121.73$ m~s$^{-1}$, $P$ is
scaled by $P_{CJ} = 17.12$ atm, time is scaled by the eddy turnover time
$\tau_{ed} = 0.367$ ms. Vertical dashed lines indicate times corresponding to
frames shown in Fig.~\ref{f:CH4}. \textit{Panel D:} $S_T$ and maximum
instantaneous $P$ in the computational domain in the simulation of a
thermonuclear turbulent flame in pure $^{12}$C at initial density $\rho_0 =
4\times10^8$ g~cm$^{-3}$. $S_T$ is scaled by $S_{CJ} = 4.8\times10^8$ \cms, $P$
is scaled by $P_{CJ} = 3.64\times10^{26}$ ergs~cm$^{-3}$, time is scaled by the
\revi{integral-scale eddy turnover time $\tau_{ed} = 5.76\times10^{-11}$ s}.
Vertical dashed lines indicate times corresponding to frames shown in
Fig.~\ref{f:thermo}. Experimental uncertainties in  quantities shown in panels A
and B are indicated with error bars \cite{Suppl}.
}
\label{f:graphs}
\end{figure}

Once the flame speed exceeds $S_{CJ}$, transition to a detonation occurs as a
catastrophic runaway process, which results in a rapid pressure build-up and
creates strong shocks inside the turbulent flame. The evolution of a chemical
turbulent flame in this process and its ultimate transition to a detonation are
shown in Fig.~\ref{f:CH4} $\&$ Movie S1 (this simulation was previously reported
elsewhere ~\cite{Poludnenko_2011}). The corresponding histories of the turbulent
flame speed and maximum pressures in the domain are shown in
Fig.~\ref{f:graphs}C.

This condition stems from the fact that $S_{CJ}$ is the maximum possible speed
of a steady-state deflagration, which satisfies conservation laws
\cite{Williams_1985}. The actual flame speed $S_T$ depends on the turbulent
intensity and the turbulent-flame structure, and can exceed $S_{CJ}$. Because
conservation laws prohibit a steady-state deflagration wave for $S_T > S_{CJ}$,
the flow would evolve producing shocks and accelerating, potentially reaching a
new steady state permitted by the conservation laws. As a result, the flame
could either become pulsatingly unstable \cite{Poludnenko_2015} periodically
oscillating between the speed above and below $S_{CJ}$ and producing shocks, or
it could transition to a CJ detonation, which can also be viewed as a CJ
deflagration coupled to a shock.

Equation~\ref{e:SCJ} is equivalent to \cite{Poludnenko_2011}
\beq
\dot e \gtrsim e/t_s,
\label{e:SCJ2}
\eeq
where $e$ is the internal (thermal) energy, $\dot e$ is the energy release rate
in a flame, and $t_s = \delta_T / c_s$ is the sound crossing time of a turbulent
flame with width $\delta_T$. Equation~\ref{e:SCJ2} essentially provides
the physical meaning for Equation~\ref{e:SCJ}: when the flame speed exceeds the
CJ threshold, burning releases within a sound-crossing time the amount of energy
close to, or greater than, the internal energy stored in the flame volume.
This results in a build-up of pressure, and ultimately causes the formation of
strong shocks.

The criterion defined by Equation~\ref{e:SCJ} does not predict the onset of tDDT
in a given turbulent reactive flow. Instead, we need to determine the
corresponding critical turbulent conditions, at which the turbulent flame speed
can reach $S_{CJ}$ and thus allow the turbulence-induced pressure runaway to
occur.

The tDDT process can occur while the flame remains in the flamelet regime,
defined as the combustion regime in which the turbulent flame can be viewed as a
folded laminar flame sheet with the internal structure minimally affected by
turbulence \cite{Poludnenko_2011}. Other theoretical models
\cite{Khokhlov_1997b, Niemeyer_1997, Woosley_2007, Ciaraldi_2013} have suggested
that turbulence-driven DDT requires formation of distributed flames, and thus
necessitates higher turbulent intensities capable of disrupting the internal
flame structure. However, the tDDT mechanism, which triggers the pressure
runaway and ultimately DDT, is not dependent on the structure of a turbulent
flame or the particular combustion regime as long as Equation~\ref{e:SCJ} is
satisfied. This process can be observed even in idealized one-dimensional flames
by artificially increasing the flame speed \cite{Gamezo_2011}. Therefore, a
similar set of critical turbulent conditions can be obtained also for
distributed flames. We focus on the derivation of the critical turbulent
conditions in the flamelet regime.

In the flamelet regime, the turbulent flame speed $S_T$ is \cite{Driscoll_2008}
\beq
S_T = I_M S_L \frac{A_T}{L^2},
\label{e1}
\eeq
where $A_T$ is the surface area of the flame sheet folded in a volume of size
$L$, $S_L$ is the laminar flame speed, and the coefficient $I_M$ accounts for
the effects of turbulent stretch. For thermonuclear flames, $I_M$ is of order
unity \cite{Dursi_2003}.

In the volume $L^3$, turbulence would fold the flame on all scales greater than
some small scale $\lambda_f$, on which the turbulent intensity $U_{\lambda}
\equiv U(\lambda)$ can be estimated as
\beq
U_{\lambda} = \alpha I_M S_L.
\label{e:Ulam}
\eeq
As a result of such packing, the flame surface density in a unit volume is equal to
the inverse of the average flame-sheet separation, which is effectively the
scale of the smallest flame folds $\lambda_f$
\beq
\frac{A_T}{L^3} = \frac{1}{\lambda_f}.
\label{e2}
\eeq
Based on the criterion in Equation~\ref{e:SCJ}, at the onset of the runaway,
\reviii{$S_T = c_s/\alpha$}. Substituting this along with Equation~\ref{e2} into
Equation~\ref{e1}, we can find the characteristic flame volume $L_{CJ}$, in which
the flame can achieve the CJ deflagration conditions
\beq
L_{CJ} = \lambda_f \frac{c_s}{\alpha I_M S_L}.
\label{e:LCJ}
\eeq
In particular, the smallest possible size of the flame region, in which tDDT can
occur, $L_{CJ}^{\mathrm{min}}$, corresponds to the maximally tight flame
packing, in which the average separation of flame-sheets, $\lambda_f$, is close
to the laminar flame thickness, $\delta_L$,
\beq
L_{CJ}^{\mathrm{min}} = \delta_L \frac{c_s}{\alpha I_M S_L}.
\label{e:LCJmin}
\eeq

We assume that turbulent properties of the reacting flow field can be
effectively represented by homogeneous isotropic Kolmogorov-type turbulence. In
reality, the presence of exothermic reactions can change the turbulence
structure inside the flame \cite{Hamlington_2011, Hamlington_2012, Towery_2016}.
These changes are primarily driven by the fluid expansion as well as the
increase of the temperature-dependent viscosity, and such effects can be
pronounced in chemical flames. In contrast, in thermonuclear flames, their
impact is minimal due to very low density ratios across the flame $\alpha
\lesssim 2$ and extremely small ratios of viscosity to thermal conduction in
degenerate plasmas defined via the Prandtl number $Pr \sim 10^{-5}$
\cite{Timmes_Woosley_1992}.

Using Kolmogorov scaling, we can find turbulent intensity $U_l^{\mathrm{max}}$
at a characteristic integral scale, $l$, which would produce such maximally
tight flame packing corresponding to Equation~\ref{e:LCJmin}
\beq
U_l^{\mathrm{max}} = \alpha I_M S_L \bigg(\frac{l}{\delta_L}\bigg)^{1/3}.
\label{e:Ulmax}
\eeq
Here we used Equation~\ref{e:Ulam} as the turbulent intensity at the scale
$\delta_L$. This integral turbulent velocity would provide the maximum burning
rate per unit volume as it would create the most tightly packed flame
configuration. Therefore, it would create conditions for the onset of tDDT in
the smallest possible volume.

At the same time, CJ conditions can also be realized at lower turbulent
intensities $U_l < U_l^{\mathrm{max}}$. In this case, the smallest flame folds
will occur on scales $\lambda_f > \delta_L$, and the resulting critical flame
volume $L_{CJ}$ required to achieve the CJ burning speed (Equation~\ref{e:LCJ})
would be larger than $L_{CJ}^{\mathrm{min}}$. In particular, turbulent intensity
$U_{CJ}$ at the scale $L_{CJ}$, which would provide sufficient flame packing on
that scale to achieve $S_T = S_{CJ}$, can be found as
\beq
U_{CJ} = U_{\lambda}\bigg(\frac{L_{CJ}}{\lambda_f}\bigg)^{1/3}.
\label{e3}
\eeq
Using Equation~\ref{e:Ulam} and~\ref{e:LCJ}, this gives
\beq
U_{CJ} = \big(\alpha I_M S_L)^{2/3}c_s^{1/3}.
\label{e:UCJ}
\eeq
This turbulent intensity is independent of scale and only depends on the
properties of the reactive mixture. Therefore, we can use Equation~\ref{e:UCJ}
and the Kolmogorov scaling to express $L_{CJ}$ in terms of the characteristic
turbulent integral scale, $l$, and velocity, $U_l$, in a given system
\beq
L_{CJ} = \big(\alpha I_M S_L\big)^2 c_s \frac{l}{U_l^3}.
\label{e:LCJ2}
\eeq

The critical flame speed $S_T = S_{CJ}$ at the scale $L_{CJ}$ can be, and
typically is, higher than the turbulent velocity $U_l=U_{CJ}$ at that scale (see
Fig.~\ref{f:graphs}). Thus, such $S_T$ does not represent the burning speed of a
flame in equilibrium with the ambient turbulent flow field. For $S_T > U_l$, the
upstream turbulence cannot supply fresh fuel to the turbulent flame sufficiently
fast to sustain such high burning rates, and faster burning can be supported
only by the fuel already contained within the flame. Such unsteady
configurations can form either when a flame enters the region of sufficiently
fast turbulence, or when turbulent intensity gradually increases above the
critical values given by $U_{CJ}$ or $U_l^{\mathrm{max}}$. Ultimately, this
unsteady burning continues as a runaway process either until a detonation is
formed, or the fuel in the flame burns out, which can lead to the pulsating
instability of turbulent flames associated with the periodic formation of shocks
\cite{Poludnenko_2015}.

Equations~\ref{e:LCJmin}, \ref{e:Ulmax}, \ref{e:UCJ}, and \ref{e:LCJ2} provide
the criteria for the onset of the pressure runaway, at which $S_T = S_{CJ}$. In
this sense, they represent only the necessary conditions for the detonation
initiation. The sufficiency condition, which defines whether a shock or a
detonation form as a result of the pressure runaway, depends on the duration of
this process. If the fuel in the turbulent flame burns out within a sound
crossing time of the flame, the result is equivalent to a constant-volume
explosion, which generates pressures insufficient for a detonation initiation
(both in chemical and thermonuclear mixtures). On the other hand, if the
burn-out lasts for several sound crossing times, pressure will build up on the
upstream side of the turbulent flame due to the gradient of the sound speed in
the flame. Such gradual pressure accumulation will ultimately produce shocks of
sufficient strength to ignite a detonation.

For packed flame configurations, the ratio of the burn-out time \reviii{$t_B =
\lambda_f/(I_M S_L)$} of the flame to the sound-crossing time $t_s = L_{CJ} /
c_s$ is equal only to the flame density ratio
\beq
\frac{t_B}{t_s} = \alpha,
\label{e:tB}
\eeq
which follows from Equation~\ref{e:LCJ}. Packed configurations formed by
terrestrial chemical flames with $\alpha \sim 3-10$ \cite{Williams_1985} burn
out during several sound-crossing times. This allows for the pressure
accumulation on the upstream side of the flame and creates conditions for
shock amplification, capable of producing DDT (cf. Fig.~\ref{f:CH4}F, G, H). For
thermonuclear flames in degenerate plasmas, however, $\alpha= 1.2 - 2.0$
\cite{Timmes_Woosley_1992}, and the burn-out occurs during only one or two
sound-crossing times. The resulting shocks are similar to those produced in a
constant-volume explosion and are of insufficient strength to ignite detonations
directly. These shocks, however, can further amplify by propagating in the
density gradient of a star \cite{Charignon_2013}, or by interacting with
surrounding turbulent flames on larger scales. We thus consider
Equation~\ref{e:SCJ} as the criterion for DDT in SNIa, although the DDT
mechanism in SNIa requires an additional step for shock amplification to
DDT strengths.

These analytical results are consistent with the DNS of chemical flames
\cite{Poludnenko_2011}. $L_{CJ}^{\mathrm{min}}$ is very close to the domain size
in calculations where pressure runaway was observed \cite{Poludnenko_2011}.
Those DNS relied on a simplified physical model involving an ideal-gas equation
of state and a single-step, first-order Arrhenius chemical kinetics representing
H$_2$-air- and CH$_4$-air-like mixtures. Thus comparison with those DNS does not
determine whether this mechanism of tDDT would also be present in realistic
chemical mixtures. In order to address this issue, next we present results of an
experimental study demonstrating the tDDT process in turbulent H$_2$-air flames
propagating with super-CJ speeds.

\section*{Experimental study}
\label{Exp}

During a SNIa explosion, DDT would occur in unconfined conditions, i.e., in the
absence of walls or obstructions that could confine pressure thus creating
conditions for its build-up. It is difficult to create a similar experimental
setting, in which the evolution of a perfectly unconfined flame can be observed
from ignition to the onset of a detonation. However, even in a confined
experimental configuration, it is possible to isolate the turbulence-flame
coupling, which could lead to a pressure runaway and a DDT, from other phenomena
related to the interactions of high-speed flows with obstructions or turbulent
boundary layers. We designed an experiment to test this.

\begin{figure}[t]
\centering
\includegraphics[clip, width=1.0\textwidth]{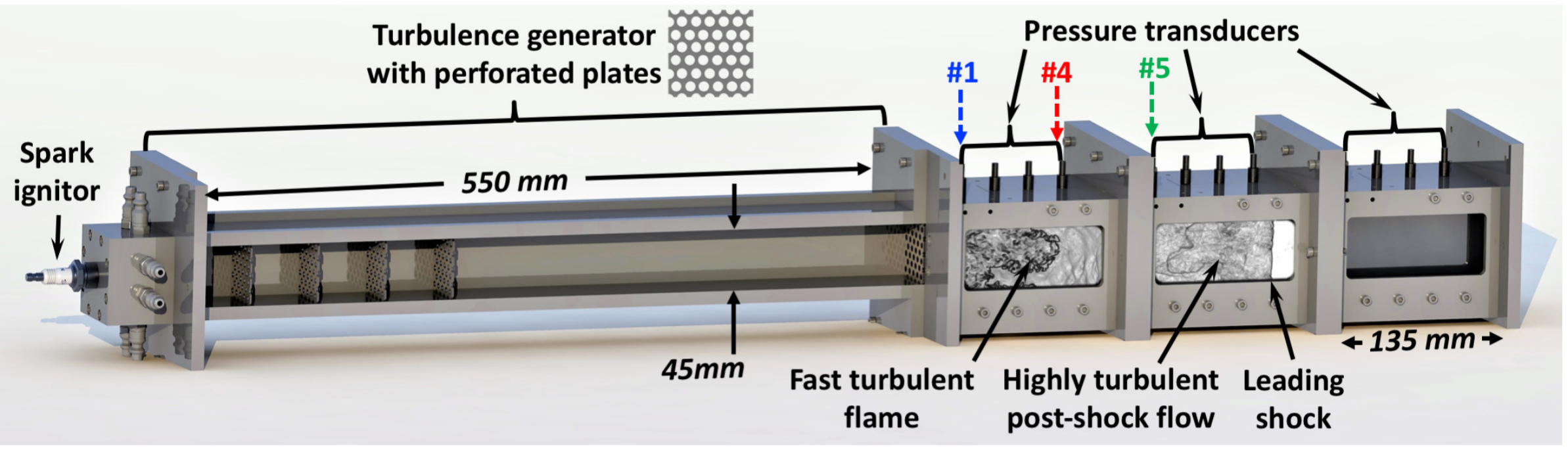}
\caption{
{\bf Experimental Turbulent Shock Tube (TST) facility.} The section with
perforated plates on the left is used to generate a fast turbulent flame, which
propagates to the right. Test section is located past the last perforated plate
at the distance of $550$ mm from the ignition point. Blue, red, and green dashed
arrows indicate the transducers $\#1$, $\#4$, and $\#5$ that produced pressure
signals shown in Fig.~\ref{f:graphs}B.}
\label{f:facility}
\end{figure}

\begin{figure}[t]
\centering
\includegraphics[clip, width=1.0\textwidth]{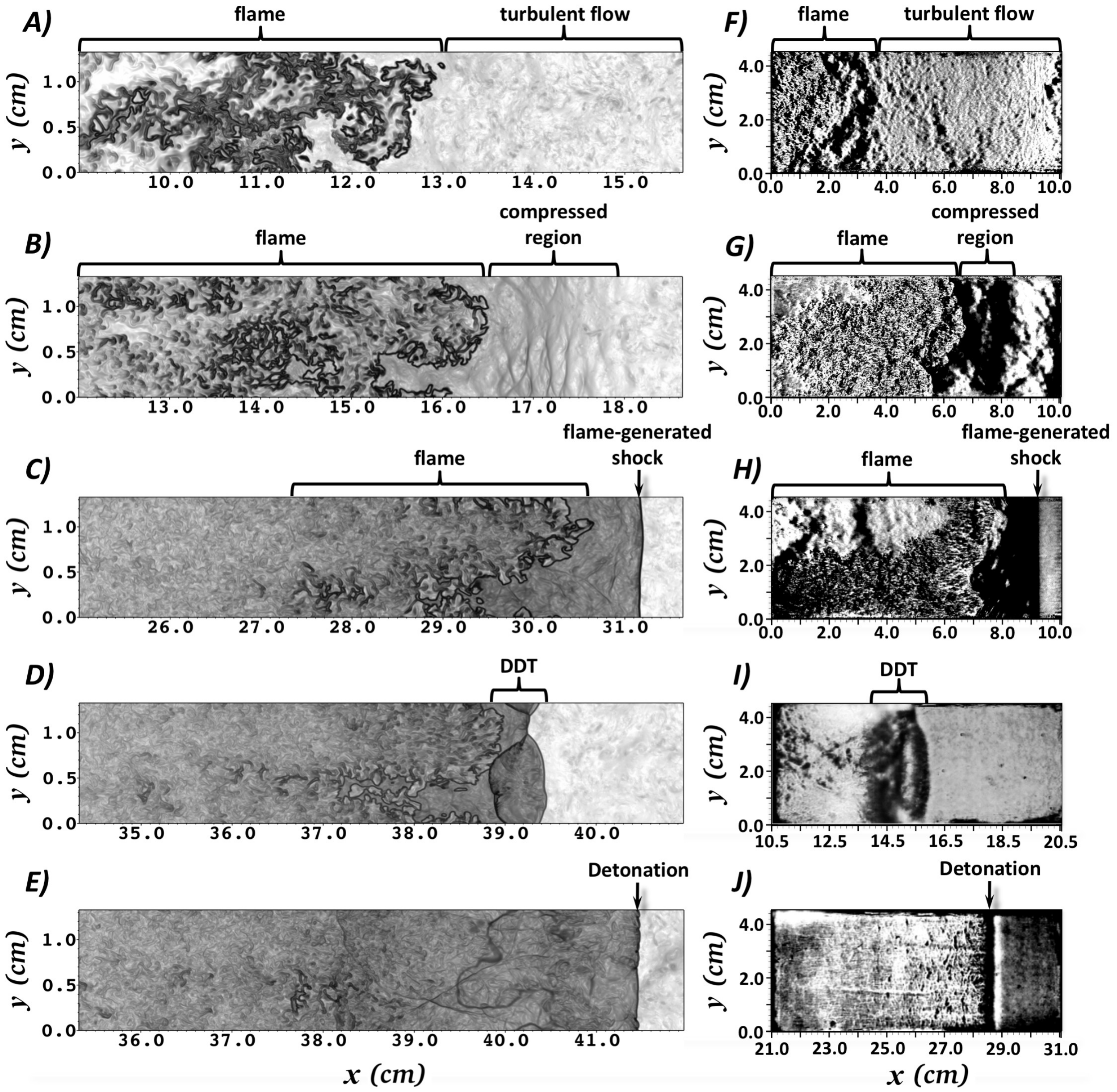}
\caption{
\reviii{{\bf Simulated (A-E) and experimental (F-J) schlieren images of the
turbulent flame during the pressure runaway and subsequent detonation
formation.} Experiment in panels F-H was carried out at $\phi=0.888$
(cf. Fig.~\ref{f:graphs}B), experiment in panels I and J was carried out
at $\phi = 0.905$. Simulated schlieren images shown in panels A-E are from the
same simulation as Fig.~\ref{f:CH4}.}
}
\label{f:schlierens}
\end{figure}

A schematic diagram of the Turbulent Shock Tube (TST) facility and the resulting
flow structure are shown in Figs.~\ref{f:facility} and \ref{f:schlierens}. This
facility is intended to create high-speed turbulent conditions, which can lead
to the spontaneous flame acceleration and tDDT that were previously modeled in
DNS \cite{Poludnenko_2011,Poludnenko_2015}. The TST consists of a $1.5$ m long
channel with one open and one closed end, with a square cross section of $45
\times 45$ mm$^2$. A spark plug is mounted at the center axis of the channel at
the closed end and used to ignite the flame. A test section is $152$ mm long
with optical access on three sides, which creates a visibility domain of
approximately $145 \times 45$ mm for advanced flame and flow-field diagnostics
\cite{Suppl}. The diagnostic section windows are composed of $25$ mm thick fused
silica designed to sustain high pressures associated with detonation waves.
Premixed hydrogen-air mixtures with varying composition are used.

Upon ignition, the flame kernel initially expands to fill the entire
cross-sectional area of the channel. Once the flame front has developed, it
begins to propagate toward the open end of the channel. A series of five
perforated plates are positioned inside the facility close to the point of
ignition to generate turbulence in the flow passing through them and thus to
produce rapid flame acceleration. As a result, a leading shock wave with Mach
$\sim 2-3$ is formed ahead of the flame. After this shock passes through the
last perforated plate immediately before the diagnostic section, the post-shock
flow creates multiple high-speed jets that produce high levels of turbulence
within the reactants. The turbulent intensity is controlled by the equivalence
ratio, $\phi$, of the initial mixture. Higher values of $\phi$ result in faster
laminar flame speeds and more rapid initial flame acceleration, which translates
into larger leading shock velocities and higher turbulence levels. The geometric
configuration employed allows us to survey the flame regimes of interest and
focus on the conditions for tDDT. Extensive experimental testing of the
design and arrangement of perforated plates has been performed to ensure that
desired turbulence conditions are achieved, specifically in terms of the
amplitude of turbulent velocity fluctuations
\cite{McGarry_2017,Chambers_2017,Sosa_2018}.

After the flame passes the last perforated plate and emerges in the diagnostic
section, it starts to interact with the high-speed turbulence in this section.
Subsequent flame evolution depends on the flame initial burning velocity, $S_T$,
which is determined by the level of turbulent fluctuations. Depending on the
value of $S_T$ relative to the CJ deflagration speed, the flame may or may not
produce strong shocks and undergo the tDDT. In this setup, varying mixture
composition over a very narrow range of $\phi$ allows us to probe the flame
dynamics, as well as the details of the tDDT process, as $S_T$ is increased from
below to above the CJ threshold.

\reviii{The turbulent flame evolution in the experiments is shown in
Fig.~\ref{f:schlierens}F-J in a sequence of schlieren images, representing flow
density gradients manifested via the gradients of the refractive index. Two
experiments are shown with mixture equivalence ratios of $\phi = 0.888$ (panels
F-H) and $\phi = 0.905$ (panels I, J) to highlight different stages of tDDT.}
Figure~\ref{f:schlierens}A-E also shows synthetic schlieren images obtained from
the DNS (see Fig.~\ref{f:CH4}) \cite{Poludnenko_2011} at the similar stages of
the flame evolution. Synthetic schlieren images represent $\log(\nabla \rho)$
computed in a 3D volume and projected onto a 2D plane.  \revi{Although an
H$_2$-air mixture was used in the experiments, the DNS shown in
Figs.~\ref{f:CH4} and \ref{f:schlierens} used a CH$_4$-air mixture, which
demonstrates the independence of the results on the choice of a reactive
mixture. The laminar flame speed in CH$_4$-air is almost an order of magnitude
lower than in H$_2$-air ($38$ \cms~vs. $302$ \cms, respectively). Calculations
using H$_2$-air mixture have been discussed elsewhere \cite{Poludnenko_2011}.}
At the time shown in Fig.~\ref{f:schlierens}F-J, boundary layers have not
developed and so their influence on the observed flame acceleration is
negligible.

\reviii{Figure~\ref{f:schlierens} shows similar dynamics in the experiment and
the DNS. Pressure waves generated within the turbulent flame propagate into the
unburned material and form a compressed region ahead of the flame
(Fig.~\ref{f:schlierens}B,G). As the runaway process develops, multiple pressure
waves coalesce into a flame-generated shock, the strength of which grows with
time (Fig.~\ref{f:schlierens}C,H). This shock forms between the flame and the
leading shock that was transmitted through the last perforated plate.
Eventually, the shock approaches the von Neumann (post-shock) pressure of a CJ
detonation (also see Fig.~\ref{f:graphs}A), at which point it triggers a DDT
(Fig.~\ref{f:schlierens}D,I) and forms a detonation
(Fig.~\ref{f:schlierens}E,J).} The case with $\phi = 0.905$
(Fig.~\ref{f:schlierens}I,J) results in a detonation, but the corresponding peak
pressure shown in Fig.~\ref{f:graphs}A is lower than $P_{CJ}$. This is because
pressures reported in Fig.~\ref{f:graphs}A were measured in the first diagnostic
window, while DDT occurred in the second window.

Figure~\ref{f:graphs}A shows peak pressures in the leading shock and those
produced by the flame, along with the corresponding turbulent flame speeds, as a
function of the equivalence ratio in multiple experiments. The evolution of the
pressure build-up inside the turbulent flame and the emergence of a
flame-generated shock for a specific experiment with $\phi=0.888$ are
demonstrated in Fig.~\ref{f:graphs}B. It provides pressure histories recorded by
several pressure transducers shown in Fig.~\ref{f:facility}.

Figure~\ref{f:graphs}A shows that the flame starts to generate strong pressure
waves when its burning velocity $S_T$ exceeds the $S_{CJ}$ threshold, i.e., when
Equation~\ref{e:SCJ} is satisfied. \revi{This results in the formation of a
compressed region of high pressure immediately ahead of the flame
(Figs.~\ref{f:schlierens}F-H). Figure~\ref{f:graphs}B shows that pressure growth
is associated with the turbulent flame and not supported by the closed end of
the channel. Pressure in the vicinity of the flame as recorded by transducers
increases by $\approx\!70\%$. This occurs as the flame propagates down the
length of the channel and passes transducers located progressively further
downstream (Fig.~\ref{f:facility}). The pressure recorded by transducer $\#1$ at
later times drops and eventually plateaus at a level well below the peak
pressures recorded in the flame region. The pressure recorded at later times by
transducer $\# 4$, which is located further ahead of transducer $\# 1$ (cf. Fig.
2), exhibits the same decreasing trend and reaches values very close to those
recorded by transducer $\# 1$. This shows that pressure upstream of the flame
front close to the last perforated plate does not change with time, which would
be the case if the observed increase in pressure in the flame region was the
result of pressurization of the entire flow from the flame to the closed end of
the channel.}

\reviii{Figure~\ref{f:graphs}A demonstrates that the turbulent flame speed in the
experiments shown in Fig.~\ref{f:schlierens} satisfies Equation~\ref{e:SCJ}. We
now consider whether turbulent conditions in these experiments are in agreement
with the theory presented above. In particular, at $\phi = 0.888$, turbulent
integral velocity, $U_l$, ahead of the flame at the beginning of the diagnostic
section, i.e., at the start of the runaway process resulting in the pressure
build-up, had an average value of $68.85$ \ms~ with the maximum and minimum
values of $238.56$ \ms~ and $9.70$ \ms, respectively, and a standard deviation of
$37.68$ \ms. Values of the turbulent integral scale, $l$, were $1.45$ cm
(average), $3.17$ cm (maximum), $0.19$ cm (minimum), and $0.45$ cm (standard
deviation). Procedures for determining these turbulent characteristics in the
TST facility have been described elsewhere~\cite{Sosa_2018}. The pressure and
temperature of the reacting mixture in the compressed region ahead of the flame
are $6.92$ atm and $627$ K, respectively. These allow us to determine the
corresponding laminar flame properties and $S_{CJ}$. The critical turbulent
integral velocity $U_l^{\mathrm{max}}$ defined in Equation~\ref{e:Ulmax} and
corresponding to the average integral scale in the flow is $166.76$ \ms. This
turbulent velocity would be required to provide maximally tight flame packing.
It is, however, larger than the average $U_l = 68.85$ \ms~ in the flow, which
implies that flame would be folded on scales larger than $\delta_L$. As a
result, the flame volume required to reach the CJ conditions would be larger
than $L_{CJ}^{\mathrm{min}}$ (Equation~\ref{e:LCJmin}) and instead would be
defined by Equation~\ref{e:LCJ}, namely $L_{CJ} = 2.6$ cm, which is within the
range of values of $l = 0.19 - 3.17$ cm observed in the experiment and $80\%$
larger than the average $l = 1.45$ cm. The corresponding $U_{CJ} = 83.64$ \ms~
(Equation~\ref{e:UCJ}) is within $20\%$, and less than $\sigma$, of the average
$U_l = 68.85$ \ms~ in the experiment. Therefore, theoretically predicted
turbulent conditions for the onset of the pressure runaway are within the range
of values observed in the experiment.}

Similar analysis can be carried out for the DNS shown in
Fig.~\ref{f:schlierens}. Turbulent integral velocity and scale in the
calculation are $U_l = 22.37$ \ms~ and $l = 0.31$ cm, respectively. Despite the
large turbulent intensities in this calculation, only the preheat zone of the
flame, where the heat and products diffuse into the cold reactants, is broadened
\cite{Hamlington_2011}. Thus, overall, combustion proceeds in the flamelet
regime, which is consistent with the theory developed above. Using the
laminar-flame properties for the single-step CH$_4$-air reaction
model~\cite{Kessler_2010}, the corresponding $U_{CJ} = 19.25$ \ms~ and $L_{CJ} =
0.2$ cm, which are just below $U_l$ and $l$ in the calculation. Therefore,
turbulent conditions in this calculation are also in agreement with the
theoretical predictions for the onset of the pressure runaway.


\section*{\revi{Numerical modeling} in thermonuclear systems}
\label{DNS}

\begin{figure}[t]
\centering
\includegraphics[clip, width=1.0\textwidth]{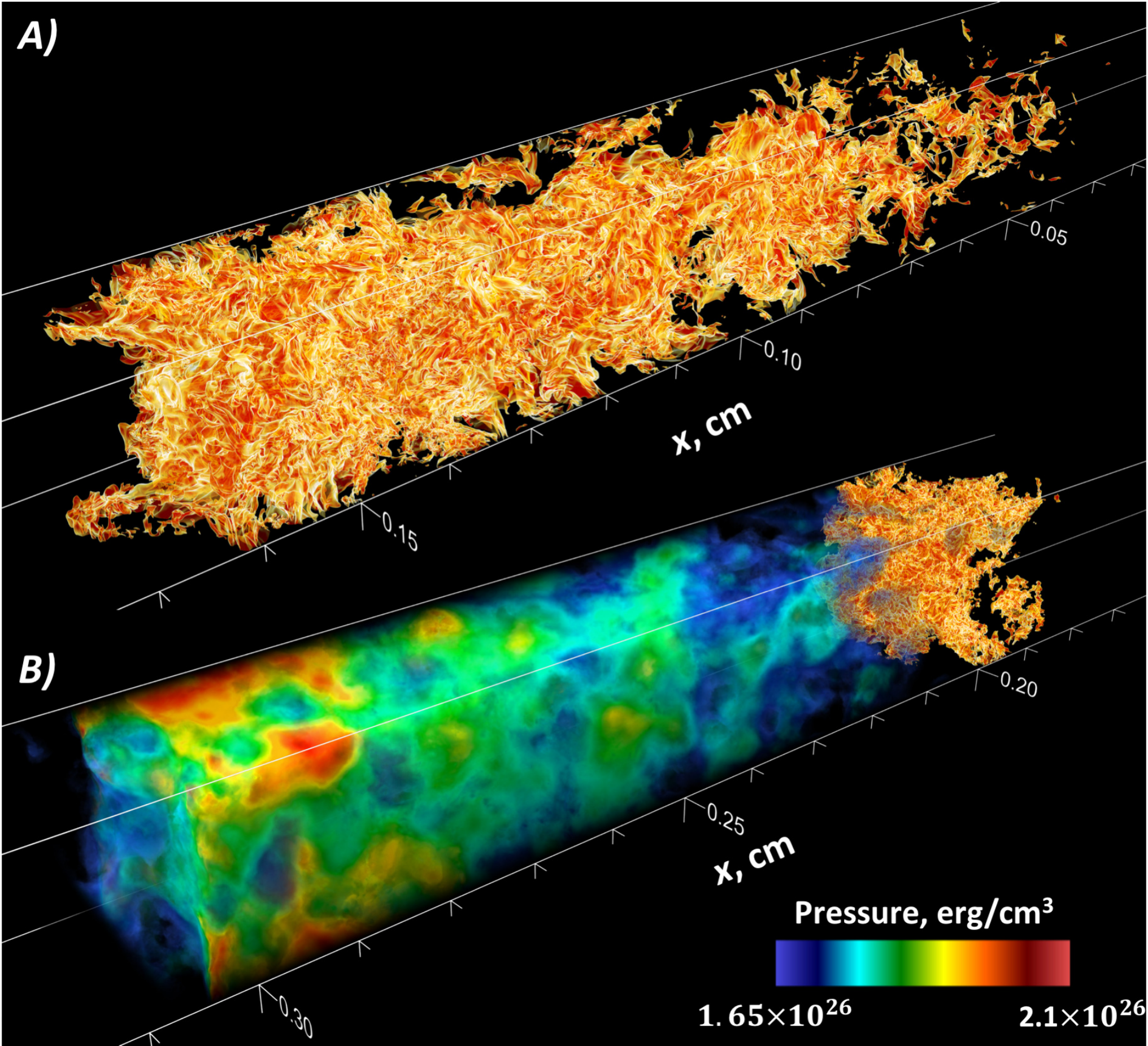}
\caption{
\textbf{Turbulence-driven spontaneous shock formation in a turbulent
thermonuclear flame.} \textit{Panel A:} flame structure at the time of maximum
turbulent burning velocity, $S_T$. \textit{Panel B:} resulting flame-generated
pressure field and the equilibrium structure of a turbulent flame at the end of
the burn-out phase. The flame propagates to the left, and is shown as a
semi-transparent flame surface with varying opacity using a ray-tracing
visualization technique. Colors correspond to the light intensity, which
increases from red to yellow to white. Corresponding turbulent flame speeds and
maximum pressures are shown in Fig.~\ref{f:graphs}D. This simulation is shown in
Movie S2.
}
\label{f:thermo}
\end{figure}

\begin{figure}[t]
\centering
\includegraphics[clip, width=0.8\textwidth]{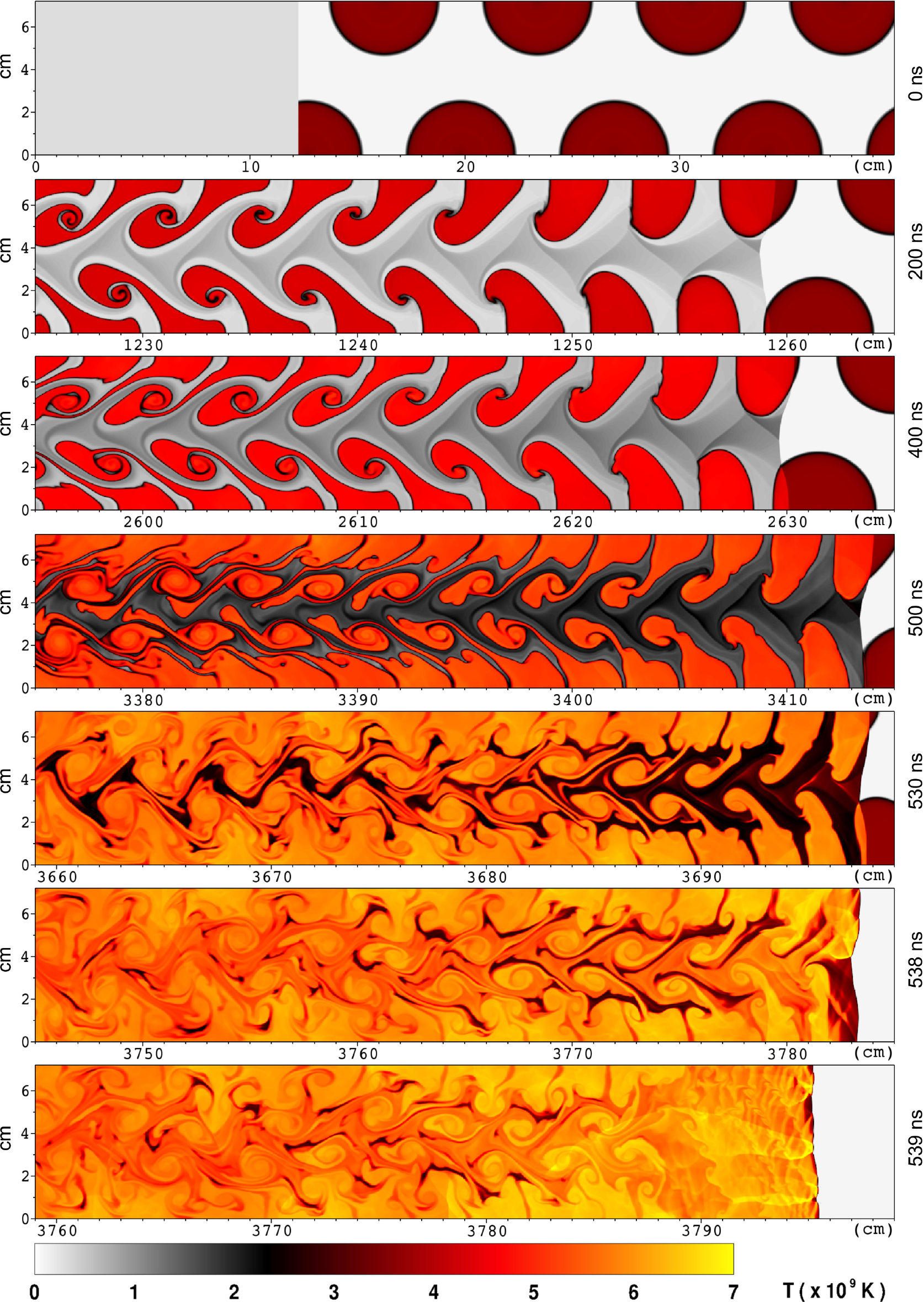}
\caption{
\revi{\textbf{Shock interaction with an idealized turbulent thermonuclear flame
and the resulting deflagration-to-detonation transition.} Shown is the
temperature distribution in a 2D calculation carried out at $\rho = 3\times
10^7$ g~cm$^{-3}$ in a $50/50$ \C/\OX~ mixture. Initial shock Mach number is
$1.28$ corresponding to a constant-volume explosion. Time since the start of the
simulation is indicated to the right of each frame. Detonation ignition occurs
at $t = 5.38\times10^{-7}$ s at $x \approx 3782$ cm.}
}
\label{f:thermo2D}
\end{figure}

\begin{figure}[t]
\centering
\includegraphics[clip, width=1.0\textwidth]{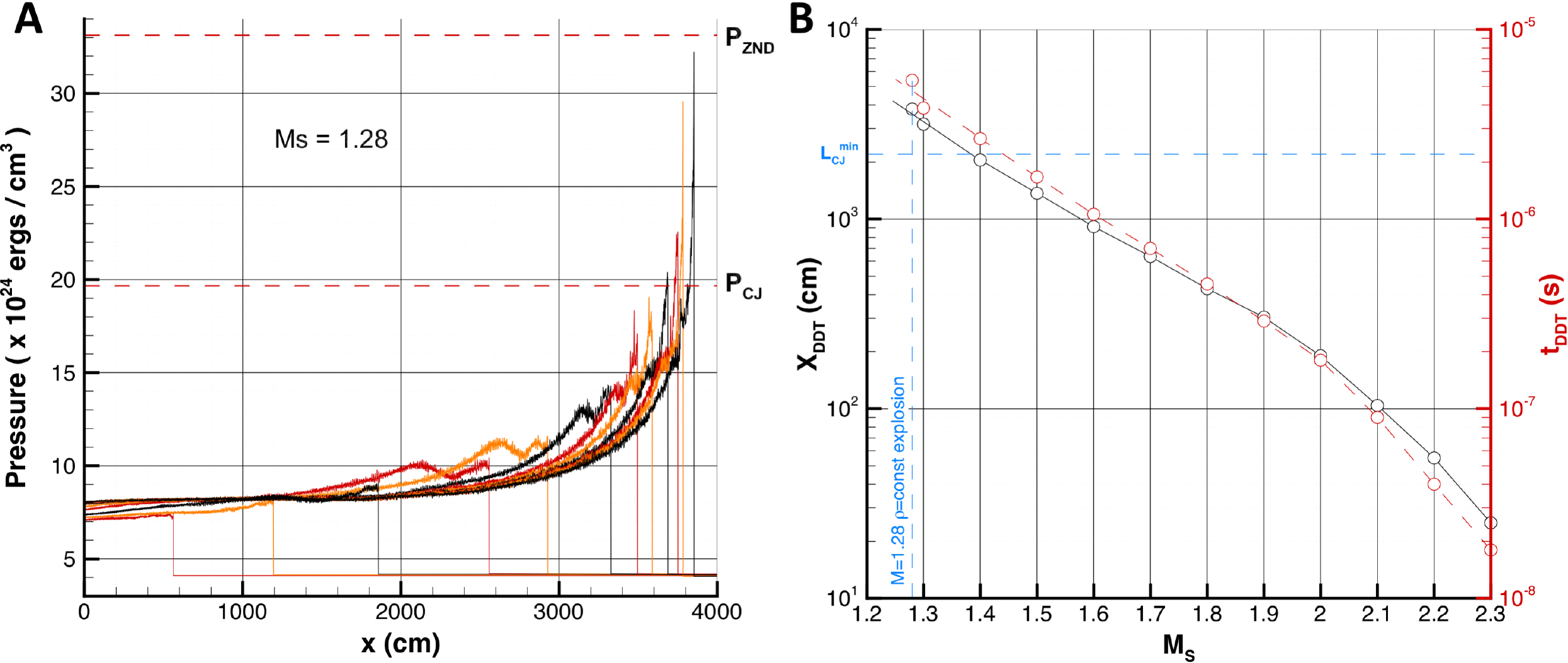}
\caption{
{\bf Deflagration-to-detonation transition driven by shock-flame interaction.}
\textit{Panel A:} Pressure evolution in the calculation shown in
Fig.~\ref{f:thermo2D}. Shown is the $y$-averaged distribution of pressure in the
domain for several time instances. Several colors are used to separate
overlapping curves. \textit{Panel B:} Distance (left axis) and time (right axis)
to DDT (cf. Fig.~\ref{f:thermo2D}) as a function of the shock Mach number.
}
\label{f:ddt}
\end{figure}

The equation of state, transport properties, and chemical kinetics that
characterize degenerate thermonuclear plasma in a WD interior during a SNIa
explosion differ in several aspects from those representative of chemical
reactive systems. Thermonuclear deflagrations are characterized by very low
density ratios, $\alpha \sim 1.2-2$ \cite{Timmes_Woosley_1992}, in contrast with
chemical flames, in which $\alpha \sim 3-10$ \cite{Williams_1985}. Because the
formation of a sonic point in a subsonic CJ deflagration is determined by the
fluid expansion in a flame, lower density ratios in thermonuclear plasmas could
affect the overall dynamics of turbulent super-CJ deflagrations. Therefore, we
next test our theoretical model under appropriate conditions of burning
thermonuclear plasmas.

Numerical simulations allow only a limited range of scales to be modeled in a
typical \revi{calculation} of turbulent flames. To capture the tDDT process in a
simulation, $L_{CJ}^{\mathrm{min}}$ for a given reactive mixture must be
sufficiently small to be accommodated in a limited computational domain.

We considered a variety of mixture compositions, from pure $^4$He and
$^4$He/$^{12}$C mixtures, which would represent sub-M$_{ch}$ and DD models, to
$^{12}$C/$^{16}$O mixtures relevant to the M$_{ch}$ scenario. We found that
$L_{CJ}^{\mathrm{min}}$ becomes sufficiently small to be simulated, namely
$L_{CJ}^{\mathrm{min}}/\delta_L \lesssim 100$, only for pure $^{12}$C mixtures
at high densities $\rho \gtrsim 2\times10^8$ g~cm$^{-3}$~\cite{Suppl}. While DDT
would not be expected to occur at such densities in a SNIa, this nevertheless
allows us to simulate tDDT in a  degenerate plasma undergoing thermonuclear
burning.

We model thermonuclear flame dynamics using compressible reactive-flow equations
solved with a finite-volume code \athena~\cite{Athena_2008, Poludnenko_2010}.
Further details of the modeling approach are given in \cite{Suppl}.

For \revi{the thermonuclear-flame simulation}, we use a flame-in-a-box
computational setup similar to the previous chemical-flame DNS
\cite{Poludnenko_2011} shown in Fig.~\ref{f:CH4}\revi{, which has also been used
in prior simulations of thermonuclear flames
\cite{Aspden_2008,Aspden_2010,Bell_2004a,Bell_2004b}}. Such calculations based
on first principles represent a small region of the flow inside a WD allowing us
to fully resolve the flame and its evolution in the turbulent flow. The flame
interacts with a homogeneous, isotropic upstream turbulence steadily driven at
the scale of the domain width using a spectral method, which introduces
divergence-free velocity fluctuations into the flow, with a prescribed energy
injection spectrum and rate \cite{Poludnenko_2010}. This approach ensures that
the turbulent integral velocity, $U_l$, and scale, $l$, in the upstream flow are
nearly constant both in space and time with a standard deviation of
$\lesssim\!2\%$ and $\lesssim\!5\%$, respectively. An analysis of the resulting
turbulence, both reacting and non-reacting, including comparison with prior
experimental and DNS results, has been presented elsewhere
\cite{Hamlington_2011, Hamlington_2012}.

The domain size and turbulent intensity were set based on the criteria given in
Equations~\ref{e:LCJmin} - \ref{e:UCJ} (see Table S1). \revi{The
calculation} uses a rectangular domain with width equal to the minimal CJ system
size $L_{CJ}^{\mathrm{min}} = 0.02$ cm (Equation~\ref{e:LCJmin}) for the fuel
density $\rho = 4\times10^8$ g~cm$^{-3}$. At this density, the laminar flame
speed $S_L = 1.35\times10^7$ \cms~ and the CJ speed $S_{CJ} = 4.8\times10^8$
\cms, or respectively $\approx\!2\%$ and $\approx\!70\%$ of the sound speed in
fuel. \revi{The calculation} is performed on a uniform Cartesian mesh with a
cell size $dx = 3.91\times10^{-5}$ cm, which corresponds to $\delta_L/dx =
4.68$. \revi{This resolution is sufficient to capture the laminar flame
properties and is similar to that used in prior studies of fully-resolved
thermonuclear deflagrations \cite{Bell_2004a,Bell_2004b}. Initially, turbulence
is allowed to evolve for $\approx\!5.2\tau_{ed}$ in order to reach an
equilibrium steady state, where $\tau_{ed} = l/U_l$ is the integral-scale eddy
turnover time. The resulting turbulence is characterized by the integral
velocity $U_l = 8\times10^7$ \cms~ $\approx\!1.84U_{CJ}$ (Equation~\ref{e:UCJ})
or $\approx\!12\%$ of the sound speed in fuel, so $U_l < S_{CJ}$. The
corresponding turbulent integral scale $l \approx L_{CJ}^{\mathrm{min}}/4$. At
the time $t = 5.2\tau_{ed}$, the thermodynamic state in the domain is
re-initialized with the exact laminar flame solution corresponding to the
initial fuel temperature $10^8$ K.} \revii{Boundary conditions are periodic in
the spanwise direction (transverse to the direction of flame propagation) and
zero-order extrapolation (outflow) along the flame-propagation direction, which
allows products to flow out from the back of the domain and does not create a
confining effect capable of promoting pressure build-up.} \revi{Despite the
substantial level of turbulent intensity, turbulent energy dissipation has a
negligible effect on the system dynamics. By the end of the calculation, fuel
temperature increases to $\approx\!8.5\times 10^8$ K from an initial $1.0\times
10^8$ K. As a result, the laminar flame speed increases by $\lesssim\!8\%$.}

Figure~\ref{f:graphs}D shows the evolution of $S_T$ and maximum pressure in the
domain in this calculation. The burning speed increases, rapidly exceeding
$S_{CJ}$ and approaching the detonation speed $D_{CJ}$. Once $S_T$ exceeds
$U_l$, however, the flame decouples from the upstream turbulence. It rapidly
consumes the fuel ingested into the flame during the early stages of the
evolution, completing the burn-out within approximately one sound crossing time,
in agreement with Equation~\ref{e:tB}. Pressure growth accelerates once $S_T$
crosses the $S_{CJ}$ threshold, similar to the behavior observed in chemical
mixtures (Fig.~\ref{f:graphs}A-C). Eventually, a strong shock forms, exits the
flame, and propagates into the fuel upstream. Figure~\ref{f:thermo} shows the
corresponding flame structure at the time of maximum $S_T$ (panel A) and after
the completion of the flame burn-out (panel B), as well as the structure of the
resulting shock wave (panel B) (also see movie S2). The resulting shock has
local peak pressures $\approx\!2\times10^{26}$ erg~cm$^{-3}$
(Fig.~\ref{f:thermo}B), and the spanwise-averaged peak pressures
$\approx\!1.85\times10^{26}$ erg~cm$^{-3}$ (cf. Fig.~\ref{f:graphs}D). The
corresponding shock Mach number $M_s \approx 1.15$. This shock is stronger than
the one produced in a constant-volume explosion at these plasma conditions, for
which the peak pressure $P_{CV} = 1.73\times 10^{26}$ erg~cm$^{-3}$ and Mach
number $M_{CV} = 1.09$ \cite{Suppl}.

Thus similar to chemical mixtures, the interaction of a highly subsonic
thermonuclear flame with a highly subsonic turbulence can produce strong shocks
in unconfined degenerate plasmas. A rapid runaway process resulting in pressure
build-up occurs once the flame speed exceeds the critical CJ deflagration
threshold in agreement with Equation~\ref{e:SCJ}. The turbulent conditions
derived above, namely, $L_{CJ}^{\mathrm{min}}$ (Equation~\ref{e:LCJmin}) and
$U_{CJ}$ (Equation~\ref{e:UCJ}), also match the onset of the runaway in
degenerate plasmas. This suggests that the overall mechanism is applicable to
both chemical and thermonuclear mixtures.

We next investigate whether shocks, which are produced in the process described
above, can ultimately trigger a detonation. Such shocks can further amplify by
propagating in the density gradient of a star \cite{Charignon_2013}, or by
interacting with surrounding turbulent flames; we focus on the latter mechanism.

While the strength of the shock observed in the DNS was larger than in a
constant-volume explosion, we assume conservatively that shocks produced by
super-CJ flames in thermonuclear plasmas are close to those in constant-volume
explosions. Propagating such a shock through a 3D turbulent flame until
detonation ignites is computationally expensive. Therefore, we demonstrate this
process in a 2D calculation, in which a turbulent flame is represented by a
series of spherical flames. An additional benefit of considering this problem in
2D is that we can model this process at a density $\rho = 3\times10^7$
g~cm$^{-3}$ and in a realistic $50/50$ \C/\OX~composition, which are closer to
the DDT conditions expected in SNIa. We consider the propagation of a
constant-volume explosion shock with Mach number $M_s = 1.28$ and
$P_{CV}=7.1\times10^{24}$ erg~cm$^{-3}$ through a series of spherical flames
with diameter $10\delta_L$ and separation between sphere centers $30\delta_L$.
Such a configuration represents a loosely packed flame. The calculation is
performed on a uniform grid with resolution $\Delta x = 0.0257$ cm or $4.4$
cells per half-\C-reaction zone length of a CJ detonation, defined as the
distance from the shock to the point at which the mass fraction of \C~ reaches
half its maximum value. Similar resolution was used in prior detonation studies,
where it was shown to reproduce the detonation velocity and resolve the
characteristic multidimensional cellular structure of unstable thermonuclear
detonations \cite{Gamezo_1999}. The size of such detonation cells is
$\approx\!3$ cm \cite{Gamezo_1999} resulting in $\approx\!2.4$ detonation cells
in the $7.2$ cm wide channel. The corresponding resolution of a laminar flame is
$9.3$ computational cells per $\delta_L$. The initial flame configuration is
shown in the upper panel of Fig.~\ref{f:thermo2D}.

As the shock begins to propagate through the turbulent flame, it compresses the
flame and generates more flame surface. Both processes accelerate burning, which
ultimately results in a pressure increase that couples to the shock and
amplifies it. This process is illustrated by the time sequence of frames in
Fig.~\ref{f:thermo2D}. Figure~\ref{f:ddt}A shows the pressure distribution
through the domain at several times. As the shock accelerates, the pressure
rapidly approaches the von Neumann value, at which point the detonation is
ignited. The resulting detonation speed is $1.16\times10^9$ \cms, which is equal
to the ideal speed of a freely propagating CJ detonation.

Figure~\ref{f:ddt}B summarizes the results of several 2D simulations performed
for the same flame configuration but different initial shock Mach numbers $M_s$.
These show the dependence of the distance and time to DDT on the initial shock
strength. Both quantities decrease rapidly with increasing $M_s$, however even
in the case of the weakest constant-volume explosion shock, the distance to DDT
is close to $L_{CJ}^{\mathrm{min}}$. Therefore, $L_{CJ}^{\mathrm{min}}$ can be
used to estimate the minimal flame region required for the tDDT process.

\section*{Transition density in the Chandresekhar-mass SNIa model}
\label{SNIa}

We use the theory developed and validated above to estimate conditions, at which
the tDDT can occur in SNIa. We restrict our analysis to the classical
Chandrasekhar-mass model.

In the M$_{ch}$ scenario, the explosion starts when a thermonuclear flame is
ignited near the WD center and propagates in the gravitational field of a WD
\cite{Gamezo_2005, Ropke_Niemeyer_2007, Ma_2013}. This flame is subject to the
Rayleigh-Taylor (RT) instability, which generates convective flows and
turbulence on multiple scales. The turbulent energy is transferred from larger
to smaller scales through a turbulent cascade. The resulting turbulent flame
propagates at subsonic speeds, thus allowing the WD to expand. The density of
the burning material changes both with the distance from the WD center and due
to the WD expansion.

Previous large-scale 3D simulations of SNIa explosions resolve scales from the
WD size $\sim\!1000$~km down to $\sim\!10$~km using a computational mesh size
$\lesssim\!1$~km \cite{Ropke_Niemeyer_2007,Ma_2013}. These scales are larger
than the characteristic scales of thermonuclear flames, which are between
$\sim\!10^{-4}$~cm and $\sim\!10$~m for $^{12}$C burning
\cite{Timmes_Woosley_1992}. Therefore, large-scale simulations rely on subgrid
models that describe the physics of flames on unresolved scales and provide the
flame speed for scales close to the computational mesh size. These simulations
do not produce shocks unless detonations are artificially triggered at some
point, and do not resolve any shock generation phenomena on scales
$\lesssim\!10$~km. Fully resolving these scales in a 3D simulation of an
exploding WD remains computationally prohibitive.

\begin{figure}[t]
\centering
\includegraphics[clip, width=1.0\textwidth]{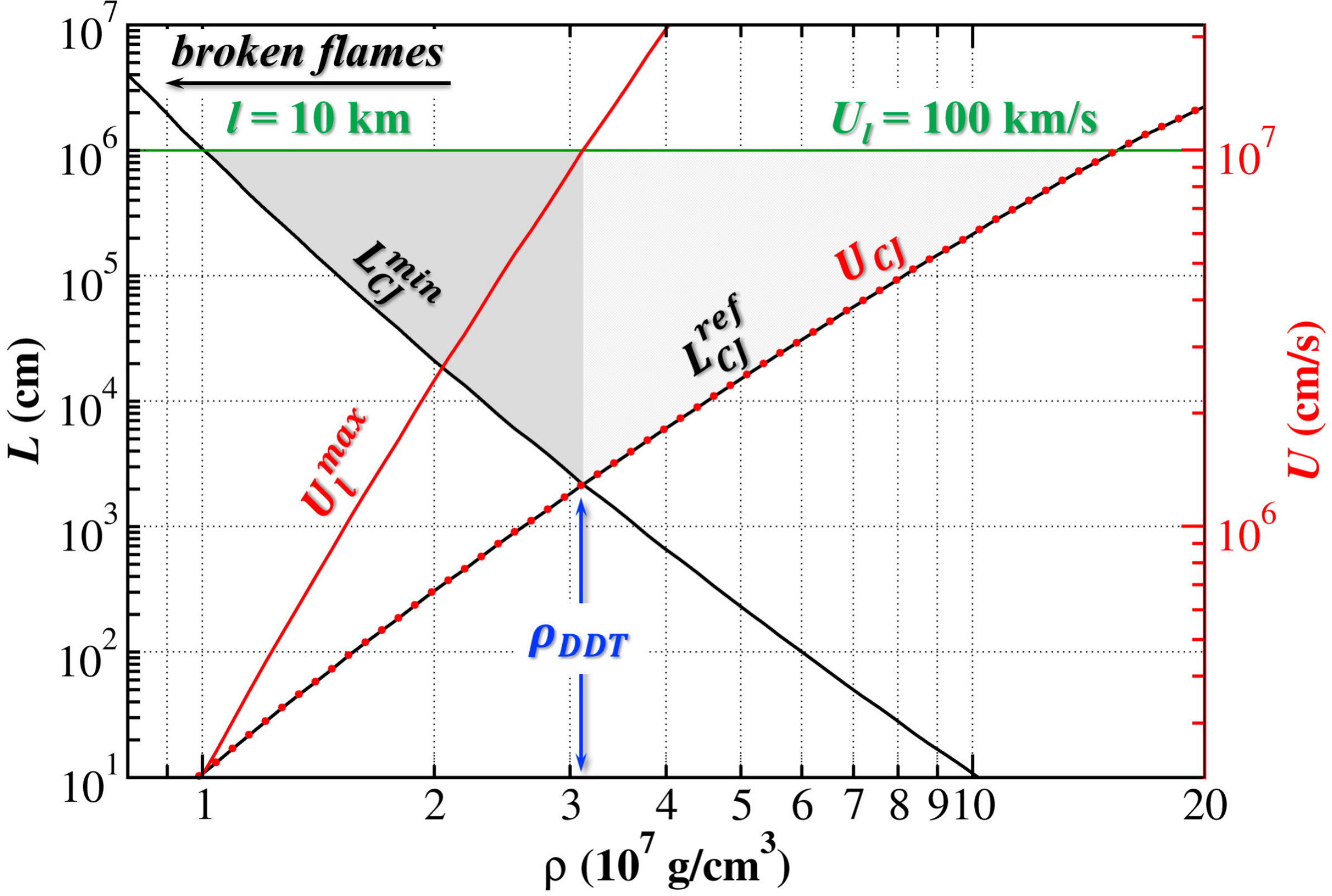}
\caption{{\bf Critical conditions for tDDT in the Chandrasekhar-mass explosion
scenario.} Quantities shown are: $L_{CJ}^{\mathrm{min}}$ (Equation~\ref{e:LCJmin}),
$U_l^{\mathrm{max}}$ (Equation~\ref{e:Ulmax}),
$L_{CJ}^{ref}$ (Equation~\ref{e:LCJ2}),
$U_{CJ}$ (Equation~\ref{e:UCJ}). 
Gray areas show the range of parameters where tDDT
is possible, with the highest probability corresponding to $\rho_{DDT}$.}
\label{f:SNIa}
\end{figure}

Using known properties of laminar thermonuclear flames in the $50/50$
$^{12}$C/$^{16}$O mixture, which represents a typical WD composition, we
computed $L_{CJ}^{\mathrm{min}}$, $L_{CJ}$, $U_l^{\mathrm{max}}$, and $U_{CJ}$
using Equations~\ref{e:LCJ}, \ref{e:LCJ2}, \ref{e:Ulmax}, and \ref{e:UCJ}. These
are shown in Fig.~\ref{f:SNIa} as functions of density. Both
$L_{CJ}^{\mathrm{min}}$ and $U_{CJ}$ depend only on the mixture properties,
which vary with local density. $L_{CJ}$ and $U_l^{\mathrm{max}}$ also depend on
the turbulent integral scale $l$ and velocity $U_l$. They were calculated
assuming $l = 10$ km and $U_l = 100$ \kms, values found in large-scale
calculations of the M$_{ch}$ explosions \cite{Niemeyer_1997, Woosley_2007,
Ropke_2007}. The lack of fully developed turbulence on scales $> 10$ km greatly
reduces flame packing on these scales.

Figure~\ref{f:SNIa} shows that $L_{CJ}^{\mathrm{min}}$ decreases below $l = 10$
km at densities $>\!10^7$ g~cm$^{-3}$, thus allowing the CJ conditions to arise
in the flow \cite{DistrFlames}. As the density increases, the minimum size of
the critical flame region decreases rapidly. At $\rho \approx 3\times10^7$
g~cm$^{-3}$, turbulent integral velocity $U_l^{\mathrm{max}}$ at the scale of
$10$ km required to produce tight flame packing associated with
$L_{CJ}^{\mathrm{min}}$ becomes larger than the reference $U_l = 100$ \kms.
Therefore, at higher densities, high turbulent intensities not observed in
full-star SNIa simulations \cite{Niemeyer_1997, Woosley_2007, Ropke_2007} would
be required to pack the flame sufficiently tightly to achieve
$L_{CJ}^{\mathrm{min}}$. CJ conditions can be created by the turbulent intensity
of $100$ \kms, though they would correspond to a less tightly packed flame and a
larger critical flame volume $L_{CJ}$. Corresponding values of $L_{CJ}$ for the
reference $U_l = 100$ \kms, that we denote $L_{CJ}^{ref}$, as well as the
turbulent velocity $U_{CJ}$ at that scale, are also shown in Fig.~\ref{f:SNIa}.
After the critical flame volume reaches a minimum value at $\rho \approx
3\times10^7$ g~cm$^{-3}$, it starts growing again, becoming larger than $10$ km
at $\rho \approx 1.5\times10^8$ g~cm$^{-3}$.

This shows that tDDT cannot occur for densities below $\approx\!10^7$
g~cm$^{-3}$ and above $\approx\!10^8$ g~cm$^{-3}$ as it would require critical
flame volumes that we do not expect to be produced by the turbulence present
during a SNIa explosion. A critical flame volume smaller than the integral scale
does not mean that tDDT will occur, given the stochastic nature of the
process. As the ratio $L_{CJ}/l$ becomes smaller, the probability of the
formation of CJ conditions within the turbulent flame increases because a smaller
size of the critical region allows for many more possible realizations in a
given flow volume of size $l$. Therefore, the maximum probability of detonation
formation is at the density corresponding to the smallest value of
$L_{CJ} \approx 2\times10^3$~cm, or $l/L_{CJ}^{\mathrm{min}} \approx 500$, at
$\rho_{DDT} \approx 3\times 10^7$ g~cm$^{-3}$. At that density, the number of
possible realizations of the CJ conditions in a volume of size $l$, which could
form over one integral-scale eddy turnover time, $\tau_l$, is
\beq N_{DDT}
\sim \Bigg(\frac{l}{L_{CJ}^{\mathrm{min}}}\Bigg)^3
\Bigg(\frac{\tau_l}{\tau_{L_{CJ}}}\Bigg) =
\Bigg(\frac{l}{L_{CJ}^{\mathrm{min}}}\Bigg)^{11/3} \sim 10^{10}.
\label{e:P}
\eeq
Here $\tau_{L_{CJ}}$ is the eddy turnover time on scale $L_{CJ}$. In other
words, the probability of the formation of a flame configuration satisfying CJ
conditions would have to be less than $10^{-10}$ to prevent the onset of the
pressure runaway at this density in a given region of size $l$. Because $l$ is 
approximately two orders of magnitude smaller than the size of the WD $\sim
1000$ km, a large number of such regions would exist during the explosion inside
the WD at the relevant densities. This further increases the probability of the
onset of the runaway, thus making tDDT in the M$_{ch}$ model almost inevitable.

\section*{Discussion and conclusions}
\label{Conclusions}

We presented a self-consistent theory of turbulence-induced shock generation and
DDT, and showed that it is in agreement with experiments involving chemical
flames and direct numerical simulations of thermonuclear deflagrations in
degenerate plasmas at conditions present in the stellar interior during a SNIa
explosion. The overall process, which triggers pressure runaway and results in
the formation of a strong shock, \reviii{qualitatively} is not sensitive to the
details of the equation-of-state, microphysical transport, or reaction kinetics.
Such runaway process will occur once the flame speed exceeds the speed of a CJ
deflagration. This theory showed that in a \C/\OX~ WD in the classical $M_{ch}$
explosion scenario, tDDT has a high probability of occurrence at densities in
the range $10^7 - 10^8$ g~cm$^{-3}$ with the maximum probability at $\rho_{DDT}
\approx 3\times10^7$ g~cm$^{-3}$. This value is similar to the transition
densities $(1.3-2.4)\times 10^7$~g~cm$^{-3}$ adopted in subgrid-scale models
used in prior large-scale calculations of the M$_{ch}$ explosion scenario
\cite{Ropke_Niemeyer_2007}, which are consistent with observations.

This analysis assumed turbulent conditions previously observed in large-scale
calculations of SNIa, namely integral velocity $U_l = 100$ \kms~ at the scale of
$l = 10$ km. This is in contrast with prior theoretical models
\cite{Khokhlov_1997a, Khokhlov_1997b, Niemeyer_1997, Woosley_2007,
Woosley_2009}, which inherently required the formation of distributed flames to
create conditions for the spontaneous reaction wave mechanism of DDT, and thus
necessitated higher turbulent intensities typically above $\sim 1000$ \kms,
which we regard as implausible.

The multiple proposed explosion scenarios \revi{for normal, bright SNIa} share a
common aspect - detonation formation at some point in the explosion. In
particular, in the context of the $M_{ch}$ scenario, our analysis suggests that
DDT is almost inevitable. The high probability of DDT given by
Equation~\ref{e:P}, however, makes it difficult for the $M_{ch}$ model to
explain the class of sub-luminous SNIa, which were previously suggested to arise
from purely deflagration-driven explosions \cite{Fink_2013}. Furthermore, the
validity of the $M_{ch}$ model has been questioned due to the lack of identified
non-degenerate companion stars surviving the explosion \cite{Shaeffer_2012} or
ejecta interaction with the companions stars \cite{Olling_2015} in some SNIa
\cite{Shappee_2018}, though observational evidence of such interaction has been
found in other events \cite{Cao_2015}. If $M_{ch}$ is not the dominant channel
for normal bright SNIa, then it becomes unclear whether WDs can grow to the
Chandrasekhar mass because once this mass limit is reached then both core
ignition and subsequent DDT would be almost unavoidable.

Our derived DDT conditions, $L_{CJ}^{\mathrm{min}}$ and $U_{CJ}$
(Equations~\ref{e:LCJmin}, \ref{e:UCJ}), depend only on the laminar flame
properties, which in turn depend on the composition of stellar material in the
interior of a WD at different radii. This suggests that the transition density
$\rho_{DDT}$ could vary substantially between 50/50 \C/$^{16}$O mixtures and
more \C-poor compositions. Because the change in $\rho_{DDT}$ would result in a
different total $^{56}$Ni yield and thus luminosity, we predict a connection  in
the $M_{ch}$ scenario between the WD age and metallicity, which determine the
interior composition, and the resulting SNIa light curve and spectral
properties. This is potentially testable with observations. There may exist a
minimal \C~mass fraction, below which the onset of DDT would become unlikely due
to either the high transition density, which would result in SNIa properties in
disagreement with observations, or turbulent conditions required to produce
super-CJ turbulent flames, which are not observed in full-star SNIa simulations.
Studies on the effect of mixture composition and metallicity on a SNIa explosion
\cite{Woosley_2007,Jackson_2010,Calder_2013} have suggested the dependence of
$\rho_{DDT}$ on the \C~ mass fraction \cite{Woosley_2007}. However, those
results were obtained for a fundamentally different DDT mechanism, which relies
on the formation of spontaneous reaction waves in a reactivity gradient produced
in a distributed flame, and an ad hoc  prescription of $\rho_{DDT}$
\cite{Jackson_2010,Calder_2013,Woosley_2007}.

\section*{Acknowledgments}
\label{Acknowledgments}

We thank the late A.M. Khokhlov (University of Chicago) for inspiring the work
on this problem. AYP and VG also thank J.C. Wheeler (University of Texas,
Austin) and E.S. Oran (Texas A\&M University) for valuable discussions.
Rendering of Figure~\ref{f:thermo} as well as movies S1 and S2 was carried out
by the Department of Defense (DoD) High Performance Computing Modernization
Program (HPCMP) Data Analysis and Assessment Center (DAAC).

{\bf Funding:} AYP and BDT were supported by the National Aeronautics and Space
Administration (NASA) award NNH12AT33I. AYP was also supported by the Air Force
Office of Scientific Research (AFOSR) award F4FGA06055G001. JC and KA were
supported by the AFOSR award FA95501610403. VG was supported by the Alpha
Foundation for the Improvement of Mine Safety and Health, Inc. through Grant
No. AFC215FO-73. Computing resources were provided by the DoD HPCMP under the
Frontier project award, and by the Naval Research Laboratory.

{\bf Author contributions:} AYP conceived and oversaw the project, developed the
theory and the code \athena, carried out DNS calculations and their analysis,
applied the theory to SNIa, and led the writing of the manuscript. JC and KA
developed the experimental facility and carried out experiments. VG developed
the concept of shock amplification to a detonation through shock-flame
interactions, contributed to the development of the overall theory and analysis
of the DNS data, application of the theory to SNIa, and carried out simulations
of shock-flame interactions. VG and KA contributed to the manuscript writing.
BDT implemented thermonuclear reaction kinetics and transport in \athena.

{\bf Competing interests:} The authors have no competing interests.

{\bf Data and materials availability:}
Experimental data for Figures~\ref{f:graphs}A,B \&~\ref{f:schlierens}F-J are
available in Data S1. Distribution of the \athena~software falls under legal
restrictions of the US government's International Traffic in Arms Regulation
(ITAR) and Export Administration Regulation (EAR). Access to the code may
require an export license from the Directorate of Defense Trade Controls, US
Department of State. Subject to that provision, readers may obtain a copy of the
software by contacting AYP. Output data for all the simulations, totaling $35$
terabytes, is permanently archived at the Department of Defense (DoD)
Supercomputing Resource Center (DSRC) at the Engineer Research and Development
Center (ERDC) in Vicksburg, Mississippi. Transfer of this large data volume can
be arranged by contacting AYP.

\subsection*{Supplementary Materials}

Materials and Methods

\noindent \revii{Figures S1-S2}

\noindent \revii{Table S1}

\noindent Movies S1-S2

\noindent Data S1

\noindent References~\cite{ThickenedFlames,Colin_2000,Gardiner_2008,Hillebrandt_2000,Caughlan_1988,Timmes_2000b,Timmes_2000a,Coquel_1998,Khokhlov_2012,Smirnov_2014,Smirnov_2015}


\clearpage


\section*{Supplementary material}

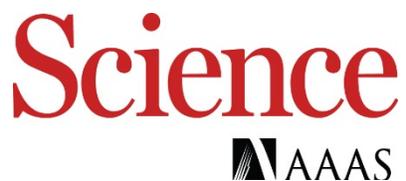

# Supplementary Materials for

## A Unified Mechanism for Unconfined Deflagration-to-Detonation Transition in Terrestrial Chemical Systems and Type Ia Supernovae

Alexei Y. Poludnenko,* Jessica Chambers, Kareem Ahmed, Vadim N. Gamezo, Brian D. Taylor

*Correspondence to: alexei.poludnenko@uconn.edu

**This PDF file includes:**

Materials and Methods
Supplementary Text
Figs. S1 and S2
Table S1
Captions for Movies S1 and S2
References *(86-96)*

**Other Supplementary Materials for this manuscript include the following:**

Movies S1 and S2
Data file S1

# Materials and methods

Here we provide further details of the experiments and numerical simulations, including the flow-field diagnostics, choice of thermonuclear plasma conditions in the 3D simulation of turbulence-flame interactions, and the characterization of the numerical simulation parameters and uncertainties.

## Experimental flow control equipment

The TST facility operation begins with the opening of the fuel and air lines and setting the appropriate flow rates for the given conditions of interest. After the fill time of 20 seconds required to ensure the uniform volumetric fill, a transistor-transistor logic signal triggers the solenoid valve to halt the chamber filling and exhaust the incoming mixture. The mixture in the chamber is allowed to settle for 3 seconds, after which a 21 ms pulse signal is sent to the ignition coil. This powers the spark plug and a flame kernel ignites. At this point, signals are sent to initiate the procedure for optical and pressure diagnostics.

The control system, shown in Fig. S1, is designed to produce a homogeneous mixture of fuel and air at standard temperature and pressure within the facility before ignition. Premixed hydrogen and air are issued into the facility at low pressures and flow rates to reduce convective velocity and initial turbulence in the channel. We used main flow rates set to achieve the desired equivalence ratio in the range $\phi = 0.78 - 1.1$. The air supplied by a compressor is regulated using an SMC Pneumatics #AW20-N02E-CZ pressure regulator and then directed to a Dwyer VFA-6-BV flow meter. Hydrogen supplied by a high pressure tank is regulated using a Specialty

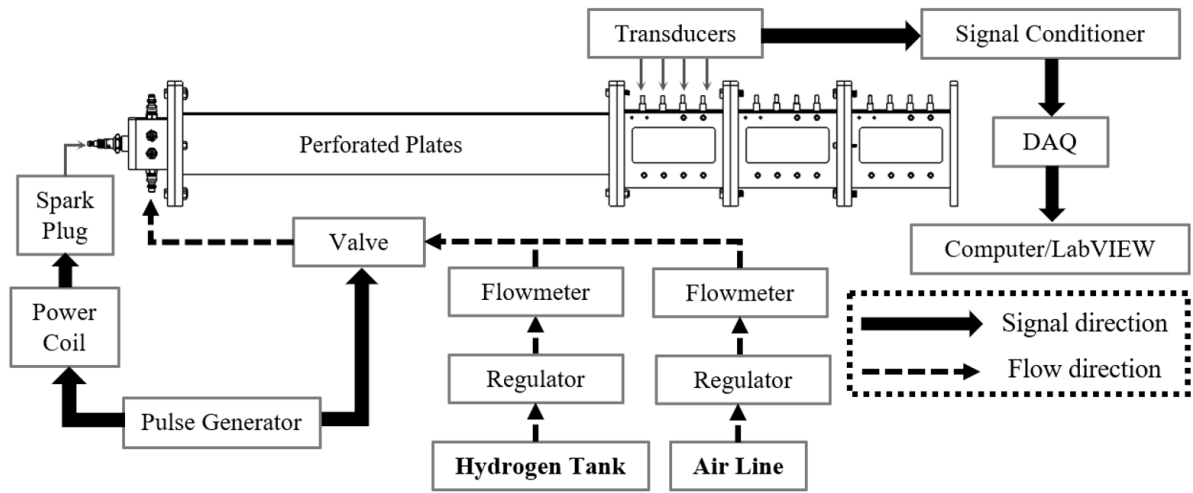

Figure S1: **Experimental setup with flow and signal direction for the TST facility shown in Fig. 3.**

Gases Southeast Inc. #HP-702-125-000-A regulator. It then flows to a VFA-3-BV flow meter for flow rate control. Desired equivalence ratios are achieved using flow rates in the range $28 - 30 \pm 0.9$ standard cubic feet per hour (SCFH) for air and $11 - 12.5 \pm 0.4$ SCFH for fuel. The fuel flow rate is corrected for use on the air-calibrated flow meter. The fuel and air lines merge to provide premixing in one line, which is then fed into the closed end of the facility. In order to prevent the flashback into the premixed line, a MAC 225B-111BAAA 3-way solenoid valve is used to divert the flow to an exhaust line immediately before ignition. A BNC Model 575 Pulse/Delay Generator is programmed with timing specifications to trigger the solenoid valve for exhaust, spark plug for ignition, and other equipment for diagnostics. The signal for the spark plug is routed to an ACCEL Super Coil 140001, which then powers the spark and immediately ignites the reactant mixture.

## Experimental diagnostics

Schlieren diagnostics are used to probe the shock and flame behavior and to quantify their respective laboratory-frame velocities throughout the test-section domain before targeting specific conditions with particle image velocimetry (PIV). A standard Z arrangement schlieren is set up using two $152.5$ mm diameter parabolic mirrors with a focal length of $1.525$ m. The images are captured using a Photron Fastcam SA-Z camera with the $1,024 \times 1,024$ pixel spatial resolution, 16-bit range depth, and $100$ kHz recording rate. The camera is equipped with Nikon lens of focal length $200$ mm and $f/2.8$. The camera captures the entire window view and the resulting image resolution is $175$ $\mu$m px$^{-1}$ with a pixel-based velocity uncertainty of $5.6$ m s$^{-1}$.

Dynamic pressure transducers are lined within the test section to capture pressure evolution in the flow. Four PCB Model #113B26 transducers are positioned with intervals of $25$ mm in the axial direction along the spanwise centerline of the top plate of the test section (Fig. 3). Transducers have sensitivity of $10$ mV psi$^{-1}$ and are operated at a frequency of $250$ kHz to resolve pressure variations in time. Considering the non-linearity and sensitivity in the transducer response, the resolved static pressure values have the uncertainty of $\pm 0.17$ atm. Transducers are connected to a PCB signal conditioner 482C Series to amplify voltage signals and then are routed to a data acquisition device with the LabVIEW control hardware and software.

Once all conditions of interest are determined using schlieren and pressure measurements, high-speed PIV and OH* chemiluminescence are used for the flow-field analysis. High-speed two-dimensional (2D) PIV is used to acquire the quantitative flow-field information for the characterization of turbulent conditions ahead of the flame. A Nd:YAG Lee LDP Dual Laser with a maximum power of $25$ mJ is operated at $532$ nm. The laser beam travels through a set of optics, which converges the beam to a diameter $< 1$ mm. The beam then reflects as a sheet up into the center of the test section to illuminate a 2D slice of the flow field. The main flow is seeded with $0.2$ $\mu$m aluminum oxide particles using a seeder designed with three levels of filtration for uniform seed density. The pulse separation of $2$ $\mu$s is intended to resolve the mean convective velocity of the shock, post-shock reactants, and the flame. The Photron Fastcam SA-Z camera equipped with a Nikon lens with focal length of $50$ mm and $f/1.2$ is

used with a frame straddling method to achieve 20 kHz PIV. Resulting PIV resolution is 42 $\mu$m px$^{-1}$ and measurement scale $\lambda_m = 168$ $\mu$m with the field of view of $44 \times 22$ mm. The imaging maintained a ratio of the particle image diameter $d_\tau$ to pixel size $d_{pix}$ at approximately $d_\tau/d_{pix} \sim 1.5$. This results in the spatial resolution of 168 $\mu$m, which is approximately a half of the laminar flame thickness, and measurement scale relative to the approximate Kolmogorov scale $\lambda_m/\eta \sim 20 - 100$. DaVis software is used for processing the PIV images with the 30 step multi-pass method leading to a $16 \times 16$ pixel interrogation box and a 75% overlap, which results in the peak velocity uncertainty $< 5$ m s$^{-1}$.

Simultaneous OH* chemiluminescence is used to capture instantaneous images of the combustion product luminescence. The flame profile can then be outlined and superimposed onto the PIV images to identify the exact flame-flow conditions. A frame rate of 40 kHz is achieved using another Photron Fastcam SA-Z camera. A Nikon lens with the focal length of 50 mm and $f/1.2$ produces a resolution of 156 $\mu$m px$^{-1}$ with the pixel-based velocity uncertainty of 3 m s$^{-1}$. In post-processing, a gamma correction of 0.465 is used to brighten the images and reveal the flame-front structure. A calibration image is taken to coordinate both the PIV and chemiluminescence domains for the flame-front alignment.

## Experimental uncertainties

The image sensor used for the flow imaging and diagnostics is a CMOS sensor with $1024 \times 1024$ pixels and $20 \times 20$ $\mu$m$^2$ pixel size. The sensor sensitivity peak is at 680 nm. The relative spectral response is 87% of peak value at 600 nm and 70% of peak value at 532 nm. The viewing test section is optically accessible on three walls, which are recessed and fitted with fused silica quartz glass. The glass allows for transmission over $250-700$ nm at 95% transmittance required for the laser diagnostics. Theoretical light transmittance in percentage, $\sigma$, of the optical system is calculated using

$$\sigma_\lambda = 100 \times \eta_{glass} \times s_\lambda, \tag{1}$$

where $\eta_{glass}$ is the relative amount of light transmitted by the optical component and $s_\lambda$ is the image sensor relative sensitivity at a given wavelength. This results in $82.65 \pm 4.1\%$ transmittance at 600 nm and $66.5 \pm 3.4\%$ transmittance at 532 nm. The uncertainty was calculated using the manufacturer's reported uncertainty and the propagation of error approach.

DaVis software is used for processing the PIV images with a 30 step multi-pass method using a $16 \times 16$ pixel interrogation and a 75% overlap resulting in a peak velocity uncertainty $< 5$ m s$^{-1}$. The flame propagation velocity, $U_f$, is estimated using a forward differencing approach

$$U_f = \frac{x_f(t + \Delta t) - x_f(t)}{\Delta t}, \tag{2}$$

where $x_f$ is the flame position.

Resulting uncertainties in all key experimentally measured quantities are shown in Fig. 2A,B and are as follows:

- Pressure: $\pm 0.17$ atm, or $\pm 0.01 P/P_{CJ}$.
- Velocity: $\pm 5$ m s$^{-1}$, or $0.026 S_T/S_{CJ}$.
- Equivalence ratio: $\pm 0.007$.

## Choice of fuel density and composition for the simulation of tDDT in thermonuclear flames

The choice of the mixture density and composition used in the numerical simulation of a turbulent thermonuclear deflagration was governed by the following considerations. The goal of this calculation is to demonstrate two points:

1. Interaction of a subsonic flame with turbulence, which is not only highly subsonic but also which has an integral velocity below $S_{CJ}$, can produce turbulent flames with burning speeds far above $S_{CJ}$.

2. When such high super-CJ burning velocities are reached, this results in the formation of strong shocks, capable of igniting a detonation.

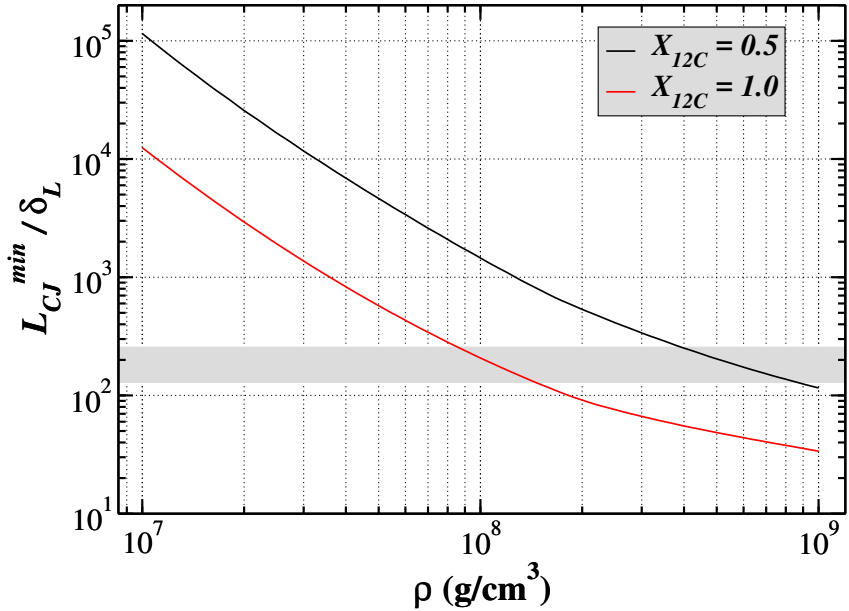

Figure S2: **Minimum size of the flame region, in which tDDT can occur, $L_{CJ}^{\min}$ (Eq. 7), normalized by the laminar flame width, $\delta_L$.** Shaded gray region shows the range of scales, which can be captured in a simulation with the domain width of $512 - 1024$ cells and the resolution $\Delta x = \delta_L/4$.

As the first step, these two effects must be demonstrated in a first-principles 3D simulation, in which the laminar flame, i.e., $^{12}$C-burning zone, is fully resolved. Carrying out such a 3D simulation, however, faces the issue of computational feasibility.

We considered a variety of mixture compositions and densities in search of the conditions, at which the tDDT could be observed. Figure S2 shows the normalized minimum turbulent flame thickness required for the tDDT, $L_{CJ}^{\min}/\delta_L$ (Eq. 7), as a function of density for two compositions: pure $^{12}$C and $50/50$ mixture of $^{12}$C and $^{16}$O. The latter composition would be more representative of the realistic white dwarf conditions, though even more $^{12}$C-poor mixtures could exist depending on the progenitor pre-explosion evolution.

In a simulation, resolution of at least $4$ cells per $\delta_L$ would be required in order to capture the flame dynamics. Consequently, assuming this flame resolution and also assuming that the domain width in a simulation is equal to $L_{CJ}^{\min}$ (which is the case in the presented calculation), Fig. S2 shows the range of scales, which can be captured in calculations with domain widths ranging from $512$ to $1024$ cells. The computational domain typically has to have an aspect ratio of at least $4:1$ to $8:1$ in the direction of flame propagation to accommodate the flame motion, though typically such aspect ratios must be much larger to allow a shock wave to develop ahead of the turbulent flame (it was $16:1$ in our final calculation). Therefore, this translates to the computational domain sizes of at least $512^3 \times 4$ to $1024^3 \times 16$, which are plausible with modern computational resources given the high computational cost of the equation of state, degenerate plasma transport, and reaction kinetics.

Figure S2 shows that demonstrating the tDDT process at densities close to $10^7$ g cm$^{-3}$, at which DDT would be expected in a $M_{ch}$ explosion scenario, would require domain sizes far beyond what is feasible with current computational resources. Figure S2 also shows that in a realistic 50/50 $^{12}$C/$^{16}$O mixture, $L_{CJ}^{\min}$ could be captured in a 512-cell-wide computational domain, which was the domain size used in our calculation, only at densities approaching $10^9$ g cm$^{-3}$. At such high densities, however, required turbulent intensities based on Eqs. 8 or 10 become so large that fuel heating due to turbulent energy dissipation would result in excessive increase in temperature. This could cause spontaneous autoignition rendering the entire flame evolution irrelevant to the DDT process. At the same time, in a pure $^{12}$C mixture, $L_{CJ}^{\min}$ can be captured in a 512-cell-wide domain at densities close to $2 \times 10^8$ g cm$^{-3}$ or above. Hence, we consider a pure $^{12}$C mixture and a slightly higher density of $4 \times 10^8$ g cm$^{-3}$.

With this choice of fuel composition and density, turbulent intensity required for the formation of a super-CJ flame was constrained based on the Eqs. 8 or 10. Although in a realistic SNIa, DDT would not be expected to occur at such turbulent intensity or in such a mixture, this choice of conditions allows one to demonstrate the two key points listed above using first-principles 3D simulations with a suitable equation of state, transport, and reaction kinetics (*86*).

## Numerical methods and uncertainties

Turbulence-flame interactions for both chemical and thermonuclear flames are modeled using the code ATHENA-RFX *(68,69)*, which is the reactive flow extension of the magnetohydro-

dynamic code Athena *(68)* (also see *(88)*). ATHENA-RFX is a fixed-grid, massively parallel, finite-volume, fully conservative code. It solves compressible reactive-flow equations, and implements a fully unsplit corner transport upwind scheme with the piecewise-parabolic (PPM) spatial reconstruction and the Harten-Lax-van Leer (HLLC) Riemann solver *(88)*. Overall, the code achieves 3rd-order accuracy in space and 2nd-order accuracy in time. A detailed description and extensive tests of the hydrodynamic integration algorithm can be found in *(68)*. Implementation of the reactive-diffusive extensions in ATHENA-RFX, along with the results of convergence studies, is discussed in detail in *(69)*.

The reaction kinetics model represents thermonuclear burning in relativistic, degenerate plasmas present in stellar interiors during SNIa explosions *(89)*. ATHENA-RFX implements a reaction network (often referred to as $\alpha$-chain) *(50)*, which includes the triple-$\alpha$, $\alpha$-capture, and heavy-ion reactions for 13 isotopes: $^4$He ($\alpha$-particle), $^{12}$C, $^{16}$O, $^{20}$Ne, $^{24}$Mg, $^{28}$Si, $^{32}$S, $^{36}$Ar, $^{40}$Ca, $^{44}$Ti, $^{48}$Cr, $^{52}$Fe, and $^{56}$Ni. Nuclear reaction rates are based on a standard tabulation *(90)* with screening corrections. The equation of state includes contributions from ideal ions, degenerate electrons, radiation, and electron-positron pairs *(91)*. Thermal conduction includes both electron and photon components with degeneracy effects *(92)*. To increase computational efficiency, both the equation of state and thermal conduction are tabulated and use bi-quadratic run-time interpolation. In thermonuclear flames, Lewis number $Le$, defined as the ratio of thermal conduction to species diffusion, $\to \infty$ and $Pr \to 0$ *(61)*. Thus, diffusion and viscosity were not included in the physical model, which makes this calculation effectively an Implicit Large-Eddy Simulation *(70-73)*.

Treatment of general equations of state in ATHENA-RFX is implemented using the energy relaxation method *(93)*. The stiff system of ordinary differential equations (ODE) for thermonuclear kinetics is integrated using a non-iterative, single-step, semi-implicit ODE integrator YASS *(94)*. YASS does not employ any approximations to the Jacobian matrix, conserves species mass fractions and total energy explicitly, and balances numerical accuracy and computational efficiency.

The convergence studies performed for chemical flames show that the accuracy of the planar laminar flame solution is $1 - 2\%$ for the resolution of 4 cells per laminar flame thickness $\delta_L$. Higher resolutions have little effect on the internal structure of flamelets, but can affect the flame wrinkling, and therefore affect the turbulent flame structure and dynamics. The turbulent flame speed decreases by $26\%$ when the computational cell size $\Delta x$ decreases from $\delta_L/8$ to $\delta_L/16$, and by another $9\%$ when $\Delta x$ decreases further to $\delta_L/32$ *(69)*. In this work, we use $\Delta x = \delta_L/8$ for chemical flames.

For thermonuclear flames, we adopt $\Delta x = \delta_L/4.68$ due to the higher computational cost of the equation of state, degenerate plasma transport, and thermonuclear reaction kinetics. This resolution allows us to reproduce the laminar flame speed with accuracy $\sim 10\%$ and is similar to the resolutions used in prior fully-resolved simulations of thermonuclear flames *(72,73)*.

We also estimated the relative error, $S_{err}$ of integration that can accumulate in 3D numerical

simulations (*95, 96*)

$$S_{err} = \sum_{i=1}^{3} (1/N_i)^{k+1} \sqrt{n}, \tag{3}$$

where $N_i$ is the number of cells in each direction, $n$ is the number of time steps, and $k$ is the order of accuracy of a numerical scheme. For the simulation parameters listed in Table S1:

- $S_{err} = 1.2 \times 10^{-7}$ for chemical flames $\quad (N = (8192, 256, 256), n = 70000, k = 3)$.

- $S_{err} = 6.5 \times 10^{-9}$ for thermonuclear flames $(N = (8192, 512, 512), n = 50000, k = 3)$.

In both cases, the estimated integration error is extremely small and does not affect the simulation results. The accuracy of the numerical solution for our simulations is controlled by the finite numerical resolution of key physical features in the flow, such as flamelets and the turbulent cascade. We therefore rely on resolution tests for accuracy estimates.

**Calculations of the constant volume explosion properties**

To compute properties of a shock produced by a constant-volume explosion, we considered the initial composition (pure $^{12}$C at $\rho = 4 \times 10^8$ g cm$^{-3}$ or $50/50$ $^{12}$C/$^{16}$O at $\rho = 3 \times 10^7$ g cm$^{-3}$) at the initial temperature $10^9$ K. We solved a zero-dimensional problem of thermonumclear burning at $\rho = const$ and adiabatic conditions by integrating the $\alpha$-network to the equilibrium composition using YASS solver. The resulting pressure was used to compute shock properties by solving the Riemann problem.

Table S1: **Simulation parameters for chemical and thermonuclear flames**

|  |  | Unit | Chemical | Thermonuclear |
|---|---|---|---|---|
| Laminar flame width | $\delta_L$ | cm | $4.154 \times 10^{-2}$ | $1.83 \times 10^{-4}$ |
| Laminar flame speed | $S_L$ | cm s$^{-1}$ | 38.024 | $1.348 \times 10^7$ |
| Fuel density | $\rho_f$ | g cm$^{-3}$ | $1.104 \times 10^{-3}$ | $4.0 \times 10^8$ |
| Fuel temprature | $T_f$ | K | 298.0 | $10^8$ |
| Fuel pressure | $P_f$ | ergs cm$^{-3}$ | $1.013 \times 10^6$ | $1.419 \times 10^{26}$ |
| Sound speed in fuel | $c_{s,f}$ | cm s$^{-1}$ | $3.314 \times 10^4$ | $6.921 \times 10^8$ |
| Domain width | $L$ | cm | 1.328 | 0.02 |
| Mesh size ($x \times y \times z$) |  | cells | $8192 \times 256 \times 256$ | $8192 \times 512 \times 512$ |
| Flame widths per $L$ | $L/\delta_L$ |  | 31.97 | 109.29 |
| Cell size | $dx$ | cm | $5.188 \times 10^{-3}$ | $3.906 \times 10^{-5}$ |
| Cells per flame width | $\delta_L/dx$ |  | 8.007 | 4.685 |
| Turbulent velocity at scale $\delta_L$ | $U_\delta$ | cm s$^{-1}$ | $1.141 \times 10^3$ | $2.696 \times 10^7$ |
| Normalized turb. velocity | $U_\delta/S_L$ |  | 30.0 | 2.0 |
| Turbulent velocity at scale $L$ | $U_L$ | cm s$^{-1}$ | $3.62 \times 10^3$ | $1.289 \times 10^8$ |
| Integral velocity | $U_l$ | cm s$^{-1}$ | $2.402 \times 10^3$ | $8.592 \times 10^7$ |
| Integral scale | $l$ | cm | 0.388 | 0.006 |
| Gibson scale | $L_G$ | cm | $1.539 \times 10^{-6}$ | $2.288 \times 10^{-5}$ |
| Inverse norm. Gibson scale | $\delta_L/L_G$ |  | $2.7 \times 10^4$ | 8.0 |
| Karlovitz number | $Ka$ |  | 164.32 | 2.883 |
| Damkohler number | $Da$ |  | 0.063 | 4.246 |
| Turbulent Mach number in fuel at scale $L$ | $M_L = U_L/c_{s,f}$ |  | 0.109 | 0.186 |
| Number of timesteps | $n$ |  | $70,000$ | $50,000$ |

**Movie S1. Evolution of a turbulent flame and transition to a detonation in a stoichiometric CH$_4$-air mixture for the same simulation as shown in Fig. 1.** The movie shows the ray-tracing-based rendering of the isosurface of the fuel mass fraction $Y = 0.5$. Colors correspond to the light intensity, which increases from red to yellow to white. The playback speed slows during the later stages of the flame evolution close to the moment of DDT.

**Movie S2. Turbulence-driven spontaneous shock formation in a turbulent thermonuclear flame, for the same simulation as shown in Fig. 5.** Shown is the isosurface of the $^{12}$C mass fraction $Y = 0.5$ along with the flame-generated pressure field. Isosurface is shown using ray-tracing-based rendering, with colors corresponding to the light intensity, which increases from red to yellow to white. Pressure is shown using ray-tracing-based volume rendering and a separate colormap given in the lower right corner. Pressure scale is in ergs cm$^{-3}$.



\end{document}